\documentclass[lettersize,journal]{IEEEtran}
\usepackage{booktabs} 
\usepackage{amsmath,amsfonts}
\usepackage{subfig}
\usepackage{algorithm}
\usepackage{array}
\usepackage{algpseudocode}%
\newcommand{\etal}{\textit{et al}. }
\usepackage{mathtools}
\usepackage{textcomp}
\usepackage{stfloats}
\usepackage{url}
\usepackage{mathtools}
\usepackage{multirow}
\usepackage{graphics}
\usepackage{enumitem}
\usepackage{CJKutf8}
\usepackage{verbatim}
\usepackage{graphicx}
\usepackage[table]{xcolor}
\usepackage{diagbox}
\usepackage{algpseudocode}
\usepackage{algpseudocode}
\usepackage{svg} 
\usepackage[utf8]{inputenc}
\usepackage{ragged2e}
\usepackage{amsmath,amssymb}

\usepackage{xcolor,cite,etoolbox}
\usepackage{hyperref}
\makeatletter 
\pretocmd\@bibitem{\csname keycolor#1\endcsname}{}{\fail}
\newcommand\citecolor[1]{\@namedef{keycolor#1}{}}
\renewcommand{\algorithmicrequire}{\textbf{Input:}}

\makeatother
\begin{document}

\title{EEGMatch: Learning with Incomplete Labels for Semi-Supervised EEG-based Cross-Subject\\ Emotion Recognition}


\author{\IEEEauthorblockN{Rushuang Zhou\textsuperscript{1},\IEEEmembership{~Graduate Student Member,~IEEE,}
Weishan Ye\textsuperscript{1},
Zhiguo Zhang\textsuperscript{1},
Yanyang Luo,
Li Zhang,\\
Linling Li,
Gan Huang,
Yining Dong,
Yuan-Ting Zhang\IEEEmembership{, Fellow,~IEEE,} and
Zhen Liang\IEEEmembership{, Member,~IEEE}}\\
\medskip

\thanks{This work was supported in part by the National Natural Science Foundation of China under Grant 62276169, 62071310, and 82272114, in part by the STI 2030-Major Projects 2021ZD0200500, in part by Shenzhen University-Lingnan University Joint Research Programme, in part by Shenzhen-Hong Kong Institute of Brain Science-Shenzhen Fundamental Research Institutions (2021SHIBS0003), in part by Shenzhen Science and Technology Research and Development Fund for Sustainable Development Project (No.KCXFZ20201221173613036), in part by Medical-Engineering Interdisciplinary Research Foundation of Shenzhen University, and in part by the InnoHK projects under the Hong Kong Center for Cerebro-cardiovascular Health Engineering (COCHE). \textit{(Corresponding author: Zhen Liang.)}}
\thanks{Rushuang Zhou, Weishan Ye, Yanyang Luo, Li Zhang, Linling Li, Gan Huang, and Zhen Liang are with the School of Biomedical Engineering, Medical School, Shenzhen University, Shenzhen 518060, China, and also with the Guangdong Provincial Key Laboratory of Biomedical Measurements and Ultrasound Imaging, Shenzhen 518060, China. (e-mail: rrushuang2-c@my.cityu.edu.hk; 2110246024@email.szu.edu.cn; 2020222077@email.szu.edu.cn; lzhang@szu.edu.cn; lilinling@szu.edu.cn; huanggan@szu.edu.cn; janezliang@szu.edu.cn).}
\thanks{Rushuang Zhou is also with Department of Biomedical Engineering, City University of Hong Kong, Hong Kong, China, and also with Hong Kong Center for Cerebro-Cardiovascular Health Engineering (COCHE), Hong Kong Science and Technology Park, Hong Kong.}
\thanks{Zhiguo Zhang is with Institute of Computing and Intelligence, Harbin Institute of Technology, Shenzhen, China, and also with Peng Cheng Laboratory, Shenzhen, China. (e-mail: zhiguozhang@hit.edu.cn).}
\thanks{Yining Dong is with School of Data Science, City University of Hong Kong, Hong Kong, and also with Hong Kong Center for Cerebro-Cardiovascular Health Engineering (COCHE), Hong Kong Science and Technology Park, Hong Kong. (e-mail: yinidong@cityu.edu.hk).}
\thanks{Yuan-Ting Zhang is with Department of Biomedical Engineering, City University of Hong Kong, Hong Kong, China, and also with Hong Kong Center for Cerebro-Cardiovascular Health Engineering (COCHE), Hong Kong Science and Technology Park, Hong Kong. (e-mail: yt.zhang@cityu.edu.hk).}

}

\markboth{Journal of \LaTeX\ Class Files,~Vol.~14, No.~8, August~2021}%
{Zhou \MakeLowercase{\textit{et al.}}: EEGMatch: Learning with Incomplete Labels for Semi-Supervised EEG-based Cross-Subject Emotion Recognition}


\maketitle

\begin{abstract}
Electroencephalography (EEG) is an objective tool for emotion recognition and shows promising performance. However, the label scarcity problem is a main challenge in this field, which limits the wide application of EEG-based emotion recognition. In this paper, we propose a novel semi-supervised transfer learning framework (EEGMatch) to leverage both labeled and unlabeled EEG data. First, an EEG-Mixup based data augmentation method is developed to generate more valid samples for model learning. Second, a semi-supervised two-step pairwise learning method is proposed to bridge prototype-wise and instance-wise pairwise learning, where the prototype-wise pairwise learning measures the global relationship between EEG data and the prototypical representation of each emotion class and the instance-wise pairwise learning captures the local intrinsic relationship among EEG data. Third, a semi-supervised multi-domain adaptation is introduced to align the data representation among multiple domains (labeled source domain, unlabeled source domain, and target domain), where the distribution mismatch is alleviated. Extensive experiments are conducted on two benchmark databases (SEED and SEED-IV) under a cross-subject leave-one-subject-out cross-validation evaluation protocol. The results show the proposed EEGMatch performs better than the state-of-the-art methods under different incomplete label conditions (with 6.89\% improvement on SEED and 1.44\% improvement on SEED-IV), which demonstrates the effectiveness of the proposed EEGMatch in dealing with the label scarcity problem in emotion recognition using EEG signals. The source code is available at \textit{https://github.com/KAZABANA/EEGMatch}.
\end{abstract}

\begin{IEEEkeywords}
Electroencephalography (EEG); Emotion Recognition; Semi-Supervised Learning; Multi-Domain Adaptation; Transfer Learning.
\end{IEEEkeywords}

\footnotetext[1]{\hspace{1mm}These authors contributed equally to this work.}

\section{Introduction}
\label{sec:introduction}
\IEEEPARstart{E}{motion} recognition using electroencephalography (EEG) signals is a rapid and exciting growing field that recently has attracted considerable interest in affective computing and brain-computer interface \cite{si2023cross}. The majority of the existing EEG-based emotion recognition studies are based on traditional supervised learning, in which all the data in the training set are well labeled. Unfortunately, labeled data are typically difficult, expensive, and time-consuming to obtain, requiring a tremendous human effort. On the other hand, unlabeled data are easier to collect in contrast to labeled data. Compared with supervised learning, semi-supervised learning, which uses both labeled and unlabeled data, could involve less human effort, solve the modeling issue with small labeled training data, and create models that are more generalizable. However, how to design algorithms that efficiently and effectively leverage the presence of both labeled and unlabeled data for improving model learning is now a crucial challenge in semi-supervised EEG-based emotion recognition.

In the existing supervised EEG-based emotion recognition studies, transfer learning methods have been widely adopted to minimize the individual differences in EEG signals\cite{TLBCI2016,li2019regional,xu2022dagam,huang2022generator,he2022adversarial}, where the labeled data from the training subjects are considered as the \textbf{source domain} and the unknown data from the testing subjects are treated as the \textbf{target domain}. By aligning the joint distribution of the source and target domain for approximately satisfying the independent and identically distributed (IID) assumption, the individual differences in EEG signals could be alleviated and the model performance on the target domain could be enhanced \cite{li2010application, TLBCI2016}. More introductions about the existing EEG-based emotion recognition studies with transfer learning methods are presented in Section \ref{sec:relatedWorksemi}.

In order to achieve reliable model training, the current EEG-based emotion recognition studies using transfer learning methods often need a substantial amount of labeled data in the source domain. If a majority number of data in the source domain are unlabeled, the model training performance would be severely constrained. Semi-supervised learning framework provides great potential to address this problem by taking advantage of a combination of a small number of labeled data and a large amount of unlabeled data and enhancing learning tasks. For example, Zhang \etal\cite{zhang2022holistic} utilized a holistic semi-supervised learning method\cite{berthelot2019mixmatch} for EEG representation and showed the superiority of the semi-supervised learning framework when the labeled data are scarce. Further, Zhang \etal\cite{zhang2022parse} combined the holistic semi-supervised learning approaches and domain adaptation methods to improve the model performance on cross-subject emotion recognition. Still, there are three limitations in the literature, that should be thoroughly investigated and resolved in future studies on semi-supervised EEG-based cross-subject emotion recognition. \textbf{(1) Inappropriate data augmentation for EEG signals.} Previous studies utilized the Mixup method\cite{zhang2018mixup} for EEG data augmentation, which ignored the non-stationarity property\cite{rasoulzadeh2017comparative} of EEG signals. The EEG samples collected from different subjects at different trials are not IID, which contradicts the main tenet of the Mixup method. \textbf{(2) Inefficient learning framework for incomplete label task.} Previous works were based on a pointwise learning framework that relied on abundant and precise labels for model training and were impractical under label scarcity conditions\cite {bao2018classification}. \textbf{(3) Ignoring the distribution mismatch between labeled and unlabeled source data.} Previous models only considered the distribution mismatch between the labeled source data and the unknown target data\cite{zhang2022parse,berthelot2021adamatch}. The distribution alignment for the unlabeled source data was disregarded.

To address the aforementioned issues, we propose a novel semi-supervised learning framework (\textbf{EEGMatch}) for EEG-based cross-subject emotion recognition with incomplete labels. The proposed EEGMatch is composed of three main modules: EEG-Mixup based data augmentation, semi-supervised two-step pairwise learning (prototype-wise and instance-wise), and semi-supervised multi-domain adaptation. The main contributions and novelties of the present study are listed below.
\begin{itemize}
    \item \textbf{An appropriate EEG data augmentation strategy is developed.} On the basis of the defined EEG samples that meet the IID assumption, an EEG-Mixup based data augmentation method is developed to generate proper augmented samples in both the source and target domains, establishing an effective augmentation pipeline for EEG signals.
    \item \textbf{An efficient learning framework with incomplete labels is proposed.} A semi-supervised two-step pairwise learning framework, including prototype-wise and instance-wise, is proposed for proper feature learning with incomplete labels. Here, the semi-supervised prototype-wise pairwise learning explores global prototypical representation for each emotion category in a semi-supervised manner, and the semi-supervised instance-wise pairwise learning captures the local intrinsic structures of EEG samples, improving the model performance in label scarcity conditions.
    \item \textbf{A distribution alignment between labeled and unlabeled source data is incorporated.} A semi-supervised multi-domain adaptation method is designed to jointly minimize the distribution shift among the labeled source data, unlabeled source data, and unknown target data. It could significantly enhance model generalization, as demonstrated by the derived error bound.
\end{itemize}

\section{Related Work}
\label{sec:relatedWorksemi}
\subsection{EEG-based emotion recognition}
In the past ten years, the feasibility and stability of using EEG signals to identify individuals' emotional states have been widely explored\cite{De2013emotion,zheng2015investigating,li2016eegdepressive,liu2017real,zhang2018cascade,yang2019multi,khare2020adaptive,liang2021eegfusenet,jiang2023spatial}. In 2013, Duan \etal extracted EEG features to represent the characteristics associated with emotional states and used the support vector machine (SVM) for emotion classification\cite{De2013emotion}. With the rapid advancement of deep learning technologies, more and more EEG-based emotion recognition studies have been developed based on deep neural network structures. For example, Zhang \etal proposed cascade and parallel convolutional recurrent neural networks to learn effective emotion-related EEG patterns\cite{zhang2018cascade}. Song \etal introduced dynamical graph convolutional neural networks to model the multichannel EEG features and dynamically capture the intrinsic relationship among different channels\cite{song2018eeg}. To extract dynamic multilevel
spatial information among EEG channels, Ye \etal proposed a hierarchical dynamic graph convolutional network (HD-GCN)\cite{ye2022hierarchical}. Li \etal developed a multiple
emotion-related spatial model to extract discriminative graph topologies in EEG brain networks and perform efficient emotion recognition\cite{li2023effective}. Niu \etal developed a novel deep residual neural network with a combination of brain network analysis and channel-spatial attention mechanism, which enabled an effective classification of multiple emotion states\cite{niu2023brain}. Aiming at incorporating the spatial and temporal information from EEG signals, Cheng \etal developed a hybrid network (HN-DGTS) by combining temporal
self-attention and dynamic graph convolution\cite{cheng2023hybrid}. The above models are robust in subject-dependent (within-subject) emotion recognition, where the training and testing data are sampled from the same subject. However, due to the individual differences in EEG signals, the model performance would deteriorate on the subject-independent (cross-subject) emotion-recognition tasks, where the training and testing data are sampled from different subjects\cite{Zheng2016,samek2013transferring,morioka2015learning}.

\subsection{EEG-based cross-subject emotion recognition with transfer learning methods}
To tackle the individual differences in the EEG signals, the existing studies have introduced transfer learning methods to eliminate the feature distribution discrepancy extracted from different subjects and improve the model stability on EEG-based cross-subject emotion recognition tasks\cite{Zheng2016,JinDANNfisrt,LiJDA2020,he2022adversarial,huang2022generator,xu2022dagam,zhou2023pr}. Based on non-deep transfer learning methods such as transfer component analysis (TCA) and transductive parameter transfer (TPT), Zheng \etal\cite{Zheng2016} developed personalized emotion recognition models to improve the cross-subject model performance, where the proposed transfer learning framework achieved a mean accuracy of 76.31\% (the obtained mean accuracy was only 56.73\% when a traditional non-transfer learning framework was adopted). Based on deep transfer learning methods, Jin \etal\cite{JinDANNfisrt} proposed an EEG-based emotion recognition model using the domain adversarial neural network (DANN)\cite{ganin2016domain}. It was found that the models using the deep transfer learning framework outperformed those using the non-deep transfer learning methods, with the mean accuracy rising from 76.31\% to 79.19\%. Further, a series of enhanced deep transfer learning frameworks based on DANN structure has been developed for cross-subject emotion recognition. For example, Li \etal\cite{LiJDA2020} proposed a domain adaptation method to minimize the distribution shift and generalize the emotion recognition models across subjects. He \etal\cite{he2022adversarial} combined the adversarial discriminator and the temporal convolutional networks (TCNs) to further enhance the distribution matching in DANN. Huang \etal\cite{huang2022generator} developed a generator-based domain adaptation model with knowledge free mechanism to maintain the neurological information during the feature alignment. Xu \etal\cite{xu2022dagam} utilized the graph attention adversarial training and biological topology information to construct the domain adversarial graph attention model (DAGAM). Peng \etal\cite{peng2022joint} proposed a joint feature adaptation and graph adaptive label propagation method (JAGP) for performance enhancement. In comparison with the original DANN structure, these models achieved superior cross-subject emotion recognition results. In order to build reliable models, each of the aforementioned models needs a significant amount of labeled EEG data in the source domain. However, gathering enough good-quality labeled EEG data is always challenging and time-consuming. It is important to develop new transfer learning strategies for efficient cross-subject emotion recognition from a few labeled data and sufficient unlabeled data. This new setting can be named as \textbf{semi-supervised EEG-based cross-subject emotion recognition.}
 
\subsection{Semi-supervised EEG-based cross-subject emotion recognition}
In contrast to prior studies of supervised EEG-based cross-subject emotion recognition, Dan \etal\cite{dan2021possibilistic} proposed a possibilistic clustering-promoting semi-supervised model for EEG-based emotion recognition, with less requirement on the labeled training data. Due to the less consideration of the distribution shift between the source and target domains, the cross-subject emotion recognition performance was constrained. To reduce the distribution mismatch between the source and target domains, Zhang \etal\cite{zhang2022parse} proposed a semi-supervised EEG-based emotion recognition model (PARSE) based on DANN and pairwise representation alignment. The experimental results showed that the proposed method can greatly improve the semi-supervised EEG-based cross-subject emotion recognition performance.

However, there are two main limitations of the aforementioned semi-supervised EEG-based cross-subject emotion recognition studies. \textbf{(1) inappropriate selection of labeled and unlabeled data.} The previous studies randomly selected the labeled and unlabeled data in terms of the 1-second segments instead of trials. It is noted that, in the benchmark databases (SEED\cite{zheng2015investigating} and SEED-IV\cite{zheng2018emotionmeter}), one trial was composed of a number of 1-second segments, and the same emotional label was assigned to all of the 1-second segments from the same trial. Therefore, a random selection of labeled and unlabeled data in terms of 1-second segments would lead to information leaking and improper model evaluation. Considering the data continuity, we suggest the selection of labeled and unlabeled data in terms of trials to remove the effect of random seeds. Specifically, for each subject in the training set, we select the first $N$ trials as \textbf{the labeled source domain} ($\mathbb{S}$) (both EEG data and labels are available) and the remaining trials as \textbf{the unlabeled source domain} ($\mathbb{U}$) (only EEG data is available). For the subject(s) in the testing set (\textbf{the target domain} ($\mathbb{T}$)), only EEG data is available. \textbf{(2) inappropriate usage of unlabeled source data.} Both Dan \etal\cite{dan2021possibilistic} and Zhang \etal\cite{zhang2022parse} did not use the unlabeled source data for model training, which led to a significant loss of data information. Using both labeled and unlabeled source data for modeling would encourage the model to perform more reliably and generalize better to unknown data. In the present study, we propose to jointly utilize both labeled and unlabeled source data for model training and develop a semi-supervised multi-domain adaption method to address the distribution alignment issue among $\mathbb{S}$, $\mathbb{U}$, and $\mathbb{T}$.

\begin{figure*}
\begin{center}
\includegraphics[width=1\textwidth]{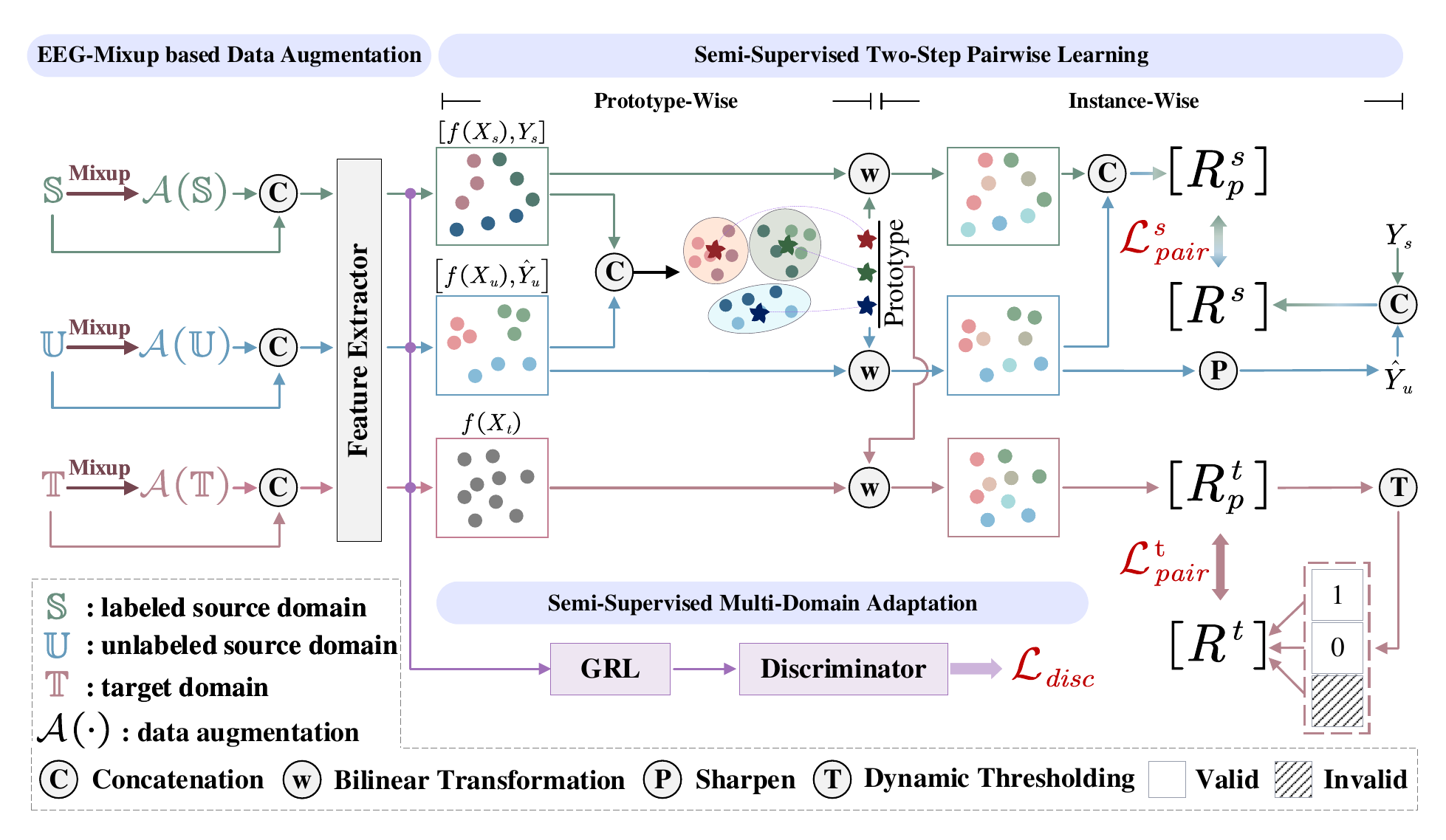}
\end{center}
\caption{The semi-supervised learning framework of the proposed EEGMatch. It consists of three modules. (1) \textbf{EEG-Mixup based data augmentation}: generate augment data and increase the sample size for modeling. (2) \textbf{Semi-supervised two-step pairwise learning}: boost the feature learning under consideration of global feature representation (prototype-wise) and local intrinsic relationship (instance-wise). (3) \textbf{Semi-supervised multi-domain adaptation}: align the distribution shift among the labeled source domain ($\mathbb{S}$), the unlabeled source domain ($\mathbb{U}$), and the target domain ($\mathbb{T}$). Here, $\mathcal{L}_{pair}^s$, $\mathcal{L}_{pair}^t$, $\mathcal{L}_{disc}$ are the semi-supervised pairwise learning loss in the source domain, the unsupervised pairwise learning loss in the target domain, and the domain discriminator loss, defined in Eq. \ref{Eq:Sourcepairwiseloss}, Eq. \ref{Eq:Targetpairwiseloss}, and Eq. \ref{Eq:Tripledomainloss}, respectively.}
\label{fig:flowchart}
\end{figure*}


\section{Methodology}
\label{sec:methodology}
Suppose the EEG samples in $\mathbb{S}$, $\mathbb{U}$, and $\mathbb{T}$ are given as ${D}_{s}=\{X_s,Y_s\}$, ${D}_{u}=\{X_u,Y_u\}$, and ${D}_{t}=\{X_t,Y_t\}$, respectively. Here, ${\{X_s,Y_s\}=\left\{\left(x_{i}^s,y_{i}^s\right)\right\}_{i=1}^{N_s}}$, ${\{X_u,Y_u\}=\left\{\left(x_{i}^u,y_{i}^u\right)\right\}_{i=1}^{N_u}}$, and ${\{X_t,Y_t\}=\left\{\left(x_{i}^t,y_{i}^t\right)\right\}_{i=1}^{N_t}}$, where $x_{i}^s$, $x_{i}^u$ and $x_{i}^t$ are EEG samples, and $y_{i}^s$, $y_{i}^u$, and $y_{i}^t$ are the corresponding emotion labels. It is noted that $y_{i}^u$ and $y_{i}^t$ are not available during model training. As shown in Fig. \ref{fig:flowchart}, the proposed model includes three main modules: EEG-Mixup based data augmentation, semi-supervised two-step pairwise learning (prototype-wise and instance-wise), and semi-supervised multi-domain adaptation, with three loss functions (semi-supervised pairwise learning loss in the source domain, unsupervised pairwise learning loss in the target domain, and multi-domain adversarial loss). \textbf{In the EEG-Mixup based data augmentation}, the augmented data, $\mathcal{A}\left(\mathbb{S}\right)$, $\mathcal{A}\left(\mathbb{U}\right)$, and $\mathcal{A}\left(\mathbb{T}\right)$, are generated in the three domains ($\mathbb{S}$, $\mathbb{U}$, and $\mathbb{T}$) separately. \textbf{In the semi-supervised two-step pairwise learning}, prototype-wise pairwise learning and instance-wise pairwise learning are involved. For the prototype-wise pairwise learning, the prototypical representation of each emotion category is learned based on the $\mathbb{S}$ and $\mathbb{U}$, and the pairwise relationships between all EEG samples and the prototypical representation of different emotion categories are measured. For the instance-wise pairwise learning, considering $Y_u$ and $Y_t$ are unknown during model training, a semi-supervised pairwise learning method is developed to reconstruct the pairwise relationship within $\mathbb{S}$ and $\mathbb{U}$, and an unsupervised pairwise learning method is adopted to reconstruct the pairwise relationships in $\mathbb{T}$. \textbf{In the semi-supervised multi-domain adaptation}, the characterized EEG features in the three domains ($f(X_s)$, $f(X_u)$, and $f(X_t)$) are trained to be as indistinguishable as possible, and the distribution shift among the three domains are aligned. More details about each module in the proposed EEGMatch are presented below.

\subsection{EEG-Mixup based data augmentation}
\label{sec:Videomixup}
The traditional Mixup method\cite{berthelot2019mixmatch} has achieved great success in the computer vision field, which requires that the pair of samples used for data augmentation should follow the joint distribution\cite{zhang2018mixup}. However, in the EEG-based emotion recognition tasks, the EEG signals collected from different subjects at different trials do not satisfy the IID assumption. In other words, a direct application of the Mixup method on the emotional EEG signals might not be appropriate.

To tackle the data augmentation problem on the EEG signals, we develop a novel EEG-Mixup method to fit the data in accordance with the IID assumption. The proposed EEG-Mixup method is mainly composed of two parts.\textbf{(1) Valid data sampling:} sampling EEG data that follows the joint distribution. Based on the prior findings\cite{Zheng2016,song2018eeg,Du2020} that the EEG data from the same trial of the same subject follow the IID assumption, the valid EEG data for data augmentation is defined as the emotional EEG signals from the same subject at the same trial, without overlapping. Take the labeled source domain ($\mathbb{S}$) as an example. The valid data sampling could be represented as $D_{p_s,q_s}^{s}=\left\{\left(x_{n}^s,y_{n}^s\right)\right\}_{n=1}^{N^s_{p_s,q_s}}$, where $N^s_{p_s,q_s}$ is the total sample size of $p_s$ subject at $q_s$ trial.
\textbf{(2) EEG data augmentation:} generating augmented data for the three domains. We conduct data augmentation on the emotional EEG signals which are randomly selected from the valid data. For example, $\left(x_i^s,y_i^s\right)$ and $\left(x_j^s,y_j^s\right)$ are the two data pairs randomly selected from $D_{p_s,q_s}^{s}$. The corresponding augmented data $\left(x_{z}^s,y_{z}^s\right)$ is computed as
\begin{equation}
    \label{Eq:mixup_x}
        x_{z}^s=\omega x_i^s+(1-\omega)x_j^s,
\end{equation}
\begin{equation}
    \label{Eq:mixup_y}
        y_{z}^s=\omega y_i^s+(1-\omega)y_j^s,
\end{equation}
where $\omega \sim \operatorname{Beta}(\alpha, \alpha)$, and $\alpha$ is a shape parameter of the $\operatorname{Beta}$ distribution. In most of the previous studies\cite{zheng2015investigating,zheng2018emotionmeter,li2019multisource,Du2020,zhong2020eeg,zhang2022parse}, the EEG samples from the same trial share the same emotional label ($y_{z}^s=y_{i}^s=y_{j}^s$). Then, the augmented data of $D_{p_s,q_s}^{s}$ could be termed as
\begin{equation}
\label{eq:mixup}
\mathcal{A}\left(D_{p_s,q_s}^{s}\right)=\left\{\left(x_{z}^s,y_{z}^s\right)\right\}_{z=1}^{Z^s_{p_s,q_s}},
\end{equation}
where $Z^s_{p_s,q_s}$ is the number of augmented samples in $\mathcal{A}\left(D_{p_s,q_s}^{s}\right)$. Similarly, the EEG-Mixup based data augmentation is also conducted in the other two domains ($\mathbb{U}$ and $\mathbb{T}$), and the corresponding augmented data are denoted as $\mathcal{A}\left(D_{p_u,q_u}^{u}\right)$ and $\mathcal{A}\left(D_{p_t,q_t}^{t}\right)$. It is noted that the emotional labels ($Y_u$ and $Y_t$) in the $\mathbb{U}$ and $\mathbb{T}$ domains are unknown during the training process, the data augmentation for $\mathbb{U}$ and $\mathbb{T}$ domains is only applied on the EEG data ($X_u$ and $X_t$), as
\begin{equation}
\mathcal{A}\left(D_{p_u,q_u}^{u}\right)=\left\{\left(x_{z}^u\right)\right\}_{z=1}^{Z^u_{p_u,q_u}}.
\end{equation}
\begin{equation}
\mathcal{A}\left(D_{p_t,q_t}^{t}\right)=\left\{\left(x_{z}^t\right)\right\}_{z=1}^{Z^t_{p_t,q_t}}.
\end{equation}
$Z^u_{p_u,q_u}$ and $Z^t_{p_t,q_t}$ are the number of augmented samples in $\mathcal{A}\left(D_{p_u,q_u}^{u}\right)$ and $\mathcal{A}\left(D_{p_t,q_t}^{t}\right)$, respectively. For simplicity, $\mathcal{A}\left(D_{p_s,q_s}^{s}\right)$, $\mathcal{A}\left(D_{p_u,q_u}^{u}\right)$, and $\mathcal{A}\left(D_{p_t,q_t}^{t}\right)$ are also represented as $\mathcal{A}\left(\mathbb{S}\right)$, $\mathcal{A}\left(\mathbb{U}\right)$, and $\mathcal{A}\left(\mathbb{T}\right)$ below. The final obtained data for the following data processing is a concatenation of the original data and the augmented data in each domain, given as  $\{\mathbb{S},\mathcal{A}\left(\mathbb{S}\right)\}$, $\{\mathbb{U},\mathcal{A}\left(\mathbb{U}\right)\}$, and $\{\mathbb{T},\mathcal{A}\left(\mathbb{T}\right)\}$.



\subsection{Semi-supervised two-step pairwise learning}
A semi-supervised two-step pairwise learning framework is proposed to enhance the feature cohesiveness of EEG samples within the same emotion class and enlarge the feature separability of EEG samples between different emotion classes.
\subsubsection{Semi-supervised prototype-wise pairwise learning}
\label{sec:Prototype}
The first step is conducted in a prototype-wise manner. The superiority of prototypical representation in feature learning has been evidenced in various research fields\cite{wang2019panet,gu2020self,Prototype2017,Pinheiro2018}. Specifically, a prototypical representation is an embedding that encodes the most representative information for a given class or relation, serving as the centroid of the features associated with the class\cite{ding2021prototypical,zhou2023pr}. The existing prototypical learning method is conducted in a supervised manner, which incorporates the label information for prototypical representation estimation. For example, for total $c$ emotion classes, the supervised prototypical representation $P$ is computed as
\begin{equation}
    \label{Eq:centroidmatrix}
    P=\left(Y^TY\right)^{-1}Y^Tf\left(X\right),
\end{equation}
where $X=[x_1;x_2,...;x_n]$ and $Y=[y_1;y_2,...;y_n]$ are the EEG data and the corresponding emotional labels (one-hot representation) in the current mini-batch, respectively. $f\left(\cdot\right)$ is a feature extractor with the parameter $\theta_f$, and $f\left(X\right)$ is a $N_B\times M$ feature matrix ($N_B$ is the batch size for stochastic gradient descent and $M$ is the feature dimensionality). $P=\left[\mu_{1};...;\mu_{i};...;\mu_{c}\right]$ and $\mu_{i}$ is the prototypical representation of the $i$th emotion class. \emph{Proof} see Supplementary Materials: Appendix A.

During the supervised prototypical learning, only the labeled samples in the source domain would be considered in the calculation of $P$. However, the sample size of the labeled source data could be very limited in most cases. To make full use of both labeled and unlabeled data in the source domain for prototypical learning, we propose a novel semi-supervised prototypical learning method by generating the corresponding pseudo labels for the unlabeled data in the source domain as
\begin{equation}
    \label{Eq:pseudo}    \hat{Y}_{u}=Sharpen\left(Cls\left(f\left(X_u\right)\right)\right),
\end{equation}
where $Cls\left(f\left(X_u\right)\right)$ is the output prediction of the emotion classifier $Cls(\cdot)$. Considering the superiority of pairwise learning against traditional pointwise learning\cite{bao2020similaritybased,khosla2020supervised,zhang2022parse,zhou2023pr}, the emotion classifier $Cls(\cdot)$ is defined to estimate the pairwise similarity as
\begin{equation}
\label{Eq:bilinear}
    l_i=w\left(f\left(x^u\right), \mu_{i}\right)=f\left(x^u\right)^{T} B \mu_{i}, i=1,...,c
\end{equation}
where $w(\cdot)$ is a bilinear transformation, $x^u$ is an EEG sample from $X_u$. $B\in R^{d\times d}$ is a trainable bilinear transformation matrix and is not restricted by symmetry or positive definiteness. $l_i$ is the calculated similarity between the data feature $f\left(x^u\right)$ and $i$th prototypical representation ($\mu_i$). Then, the predicted pseudo label for $x^u$ could be estimated as
\begin{equation}
\label{Eq:Cls}
\begin{split}
Cls\left(f\left(x^u\right)\right)&=softmax\left(\left[l_1,...,l_i,...,l_c\right]\right)\\&=softmax\left(f\left(x^u\right)^{T} B P_{s^{*}}\right).
\end{split}
\end{equation}
Further, to minimize the entropy in the predicted pseudo label, we adopt a $Sharpen$ function\cite{berthelot2019mixmatch} to adjust the “temperature” of the estimated class distribution $Cls\left(f\left(x^u\right)\right)$ as
\begin{equation}
\label{Eq:sharpen}
Sharpen\left(Cls\left(f\left(x^u\right)\right), \eta\right)_{c}=l_{c}^{\frac{1}{\eta}} / \sum_{j=1}^{C} l_{j}^{\frac{1}{\eta}},
\end{equation}
where $\eta$ is a hyperparameter for “temperature” controlling. Then, based on the labeled source data $\{X_s, Y_s\}$ and the unlabeled source data $\{X_u,\hat{Y}_{u}\}$, Eq. \ref{Eq:centroidmatrix} could be rewritten as
\begin{equation}
    \label{Eq:semicentroidmatrix}    P_{s^{*}}=\left(\left[Y_s;\hat{Y}_{u}\right]^T\left[Y_s;\hat{Y}_{u}\right]\right)^{-1}\left[Y_s;\hat{Y}_{u}\right]^Tf\left(\left[X_s;X_u\right]\right).
\end{equation}
In the implementation, for each mini-batch, $\{X_s,Y_s\}$ and $\{X_u,-\}$ are randomly sampled from $\{\mathbb{S},\mathcal{A}\left(\mathbb{S}\right)\}$ and $\{\mathbb{U},\mathcal{A}\left(\mathbb{U}\right)\}$ defined in Section \ref{sec:Videomixup}. $\hat{Y}_{u}$ is the estimated pseudo labels calculated in Eq. \ref{Eq:pseudo}. Before model training, the prototype matrix $P_{s^{*}}$ is initialized using the Eq. \ref{Eq:centroidmatrix} with only labeled source data. Note that the proposed method (Eq.\ref{Eq:semicentroidmatrix}) can be considered as a special case of supervised prototypical learning (Eq.\ref{Eq:centroidmatrix}). The main difference is that the proposed method preserves the unlabeled data for prototype computation, while the supervised method excludes them.  The samples in the target domain ($\{\mathbb{T},\mathcal{A}\left(\mathbb{T}\right)\}$) are not used in the prototypical learning to avoid overfitting.

\subsubsection{Semi-supervised instance-wise pairwise learning}
\label{sec:Instance}
The second step is conducted in an instance-wise manner. To capture the inherent relationships among the EEG data, we estimate the pairwise similarity among samples.

The instance-to-instance similarity matrix $R_p^s$ in the source domain is calculated as
\begin{equation}
\label{Eq:SourceRp}
R_p^s(i,j)=\frac{Cls\left(f\left(x^{s^*}_i\right)\right)Cls\left(f\left(x^{s^*}_j\right)\right)^{T}}{\|Cls\left(f\left(x^{s^*}_i\right)\right)\|\|Cls\left(f\left(x^{s^*}_j\right)\right)\|},
\end{equation}
where $x_{i}^{s^*}$ and $x_{j}^{s^*}$ are the EEG data sampled from $X^{s^*}=\left[X_s;X_u\right]$. $X_s \in \{\mathbb{S},\mathcal{A}\left(\mathbb{S}\right)\}$ and $X_u \in \{\mathbb{U},\mathcal{A}\left(\mathbb{U}\right)\}$. $Cls\left(f\left(\cdot\right)\right)$ is defined in Eq. \ref{Eq:Cls}. For the labeled source data $X_s$, the groundtruth is given as $Y_s$. For the unlabeled data $X_u$ in the source domain, the estimated pseudo labels are $\hat{Y}_{u}$ given in Eq. \ref{Eq:pseudo}. Then, the groundtruth similarity matrix $R^s$ in the source domain is calculated as
\begin{equation}
\label{Eq:sourceR}
R^s=N({Y^{s^*}})N({Y^{s^*}})^T,
\end{equation}
where ${Y^{s^*}}=\left[Y_s;\hat{Y}_{u}\right]$, $N(\cdot)$ is a normalization function which normalizes the column vectors of a given matrix to unit vectors. Then, the instance-wise pairwise learning loss in the source domain is defined as
\begin{equation}
\label{Eq:Sourcepairwiseloss}
\mathcal{L}_{\text{pair}}^s(\theta_f,B)=\frac{1}{N^2}\sum L^s,
\end{equation}
where $L^s=-R^s\log R_p^s-\left(1-R^s\right) \log \left(1-R_p^s\right)$. $N$ indicates the total sample size of the labeled and unlabeled data in the source domain.

For the target domain, the instance-to-instance similarity matrix $R_p^t$ is estimated as
\begin{equation}
\label{Eq:targetRp}
R_p^t(i,j)=\frac{Cls\left(f\left(x^t_i\right)\right)Cls\left(f\left(x^t_j\right)\right)^{T}}{\|Cls\left(f\left(x^t_i\right)\right)\|\|Cls\left(f\left(x^t_j\right)\right)\|},
\end{equation}
where $x_{i}^{t}$ and $x^t_j$ are sampled from the target domain data $X_t$. $Cls\left(f\left(\cdot\right)\right)$ is defined in Eq. \ref{Eq:Cls}. The corresponding groudtruth similarity matrix $R^t$ is based on a dynamic thresholding method, given as
\begin{equation}
\label{Eq:PairwiseRt}
    \begin{aligned}
 R^t\left(i,j\right)=\left\{\begin{array}{l}
    1, \text { if } R_p^t(i,j) \geq \tau_h\\
    0, \text { if } R_p^t(i,j) < \tau_l,\\
    \end{array}\right.
    \end{aligned}
\end{equation}
where $\tau_h$ and $\tau_l$ are the defined dynamic upper and lower thresholds, respectively. Then, the instance-wise pairwise learning loss in the target domain is defined as
\begin{equation}
\label{Eq:Targetpairwiseloss}
\mathcal{L}_{\text{pair}}^t(\theta_f,B)=\frac{1}{\sum_{i,j}M\left(i,j\right)}\sum_{i,j}M\left(i,j\right)L^t(i,j),
\end{equation}
where $L^t$ is a $N_t \times N_t$ pairwise loss matrix, given as $L^t=-R^t\log R_p^t-\left(1-R^t\right)\log \left(1-R_p^t\right)$.  $M$ is a $N_t \times N_t$ mask matrix to define the valid pairs involved in the loss calculation, given as
\begin{equation}
\label{Eq:Targetmask}
    \begin{aligned}
 M\left(i,j\right)=\left\{\begin{array}{l}
    1, \text { if } {R^t\left(i,j\right)=1 \text{ or } 0}  \\
    0, \text { if } {\text{otherwise}}.\\
    \end{array}\right.
    \end{aligned}
\end{equation}
$N_t$ is the total sample size of the unlabeled target domain samples $X_t$ in the current mini-batch. In other words, for the instance-to-instance similarity relationship that falls into the range of $\tau_l\leq R^t_p\left(i,j\right) <\tau_h$, the pair relationship $\left(x^t_i,x^t_j\right)$ could be considered as invalid and would be excluded in the loss calculation at this round. In order to gradually include more samples for model training, the dynamic thresholds ($\tau_h$ and $\tau_l$) are set in a progressive learning strategy, as
\begin{equation}
\label{Eq:PairTargetNonLinear}
\left\{\begin{array}{l}
\tau_h^{t}=\tau_h^{0}-\frac{\tau_h^{0}-\tau_l^{0}}{2}\times\left(1-\left(1-\frac{2}{Maxepoch}\right)^{t}\right)\\
\tau_l^{t}=\tau_l^{0}+\frac{\tau_h^{0}-\tau_l^{0}}{2}\times\left(1-\left(1-\frac{2}{Maxepoch}\right)^{t}\right),
\end{array}\right.
\end{equation}
where $\tau_h^t$ and {$\tau_l^t$} are the upper and lower thresholds at the current training epoch $t$, respectively. $Maxepoch$ is the maximum training epoch. $\tau_h^0$ and $\tau_l^0$ are the initial values for the upper and lower thresholds. 

Overall, the objective function of the instance-wise pairwise learning is a combination of the instance-wise pairwise learning loss in the source domain (Eq. \ref{Eq:Sourcepairwiseloss}) and the instance-wise pairwise learning loss in the target domain (Eq. \ref{Eq:Targetpairwiseloss}), given as

\begin{equation}
\begin{split}
\label{Eq:insancepairwiseloss}
\mathcal{L}_{\text{pair}}(\theta_f,B)  =& \mathcal{L}_{\text{pair}}^s(\theta_f,B)+\gamma\mathcal{L}_{\text{pair}}^t(\theta_f,B) \\
 & +\beta\left\|{P_{s^{*}}}^{T} {P_{s^{*}}}-{I}\right\|_{F},
\end{split}
\end{equation}
where $\left\|{P_{s^{*}}}^{T} {P_{s^{*}}}-{I}\right\|_{F}$ is a penalty term that is applied to avoid redundant feature extraction in the prototypical learning\cite{Pinheiro2018}, and $\left\|\cdot \right\|_{F}$ is the Frobenius norm. $\gamma=\delta \times \frac{epoch}{Maxepoch}$, where $\frac{epoch}{Maxepoch}$ is the ratio of the current epoch to the maximum epoch. $\beta$ and $\delta$ are the hyperparameters.

\subsection{Semi-supervised multi-domain adaptation}
\label{sec:Tripleadapt}
The traditional supervised domain adaptation methods on EEG data were conducted based on DANN\cite{ganin2016domain}, where the feature distribution in the source and target domains were aligned\cite{Du2020,2018WGAN,BiHDM2019,He2018DAN,Zhaoplug2021,zhong2020eeg}. However, in the semi-supervised learning, treating all the labeled and unlabeled source data as a single domain would increase the difficulty of adaptation\cite{li2019multisource,chen2021msmda}. To further eliminate the distribution mismatch among different domains, we introduce a novel semi-supervised multi-domain adaptation method in this paper to align the feature representation among the labeled source domain ($\mathbb{S}$), the unlabeled source domain ($\mathbb{U}$), and the target domain ($\mathbb{T}$). Through the semi-supervised multi-domain adaptation, the feature distribution discrepancy among the above three domains could be mitigated, and the generalization ability of $f(\cdot)$ and $Cls(\cdot)$ could be enhanced. The existing multi-source multi-domain adaptation methods\cite{li2019multisource,MEERNet2021,chen2021msmda} and domain generalization strategies\cite{Dresnet2019} assumed that the label information is complete in the source domain, which is not practical for real-world label scarcity applications. On the contrary, the proposed semi-supervised multi-domain adaptation method assumes that the label information is incomplete in the source domain to address the limitation of the existing methods.

The theoretical motivation of the proposed semi-supervised multi-domain adaptation method is first presented. In the existing domain adaptation methods for only two domains (labeled source domain $\mathbb{S}$ and target domain $\mathbb{T}$), the target error $\epsilon_{{T}}(h)$ could be bounded as \cite{ben2010theory}
\begin{equation}
\label{Eq:targetbound}
\epsilon_{T}(h) \leq \delta+\epsilon_{S}(h)+d_{\mathcal{H} }\left(\mathcal{D}_{S}, \mathcal{D}_{T}\right),
\end{equation}
where $\delta$ represents the difference in labeling functions across the two domains and is typically assumed to be small under the covariate shift assumption\cite{david2010impossibility}. $\epsilon_{S}(h)$ is the source error based on a given classification hypothesis $h$ characterized by the source classifier. $d_{\mathcal{H} }(\mathcal{D}_{S}, \mathcal{D}_{T})$ corresponds to the $\mathcal{H}$-divergence introduced in\cite{kifer2004detecting}, which measures the divergence between the source domain distribution $\mathcal{D}_{S}$ and the target domain distribution $\mathcal{D}_{T}$ and could be estimated directly from the error of a binary classifier trained to distinguish the domains\cite{ben2006analysis}. Alternatively, an accurate estimation of the source error and the $\mathcal{H}$-divergence is important for the optimization of the target error. 

However, in the semi-supervised learning, an accurate estimation of the source error and the $\mathcal{H}$-divergence would be extremely challenging with only a few labeled source data. To tackle this issue, we suggest using both labeled and unlabeled samples in the source domain for the estimation of the source error and the $\mathcal{H}$-divergence. In the proposed semi-supervised multi-domain adaptation method, the source domain is given as $\mathbb{S^{*}}=\{\mathbb{S},\mathbb{U}\}$, which is composed of both the labeled source data ($\mathbb{S}$) and the unlabeled source data ($\mathbb{U}$). The convex hull 
$\Lambda_{S_{\pi}^*}$ of $S_{\pi}^*$ is a set of mixture distributions, given as
\begin{equation}
    \Lambda_{S_{\pi}^*}=\{\overline{\mathcal{D}}_{S_{\pi}^*}: \overline{\mathcal{D}}_{S_{\pi}^*}(\cdot)=\pi_{S}\mathcal{D}_{S}(\cdot)+\pi_{U}\mathcal{D}_{U}(\cdot)\},
\end{equation}
where $\overline{\mathcal{D}}_{S_{\pi}^*}$ is a distribution computed by the weighted summation of the labeled source domain distribution $\mathcal{D}_{S}$ and the unlabeled source domain distribution $\mathcal{D}_{U}$. $\pi_{S}$ and $\pi_{U}$ are the corresponding weights, belonging to $\Delta_{1}$ (a 1-th dimensional simplex). For the target domain $\mathbb{T}$, $\overline{\mathcal{D}}_{T}$ is the closest element to $\mathcal{D}_{T}$ within $\Lambda_{S_{\pi}^*}$, which can be calculated as 
\begin{equation}
    \operatorname{argmin}_{\pi_{S},\pi_{U}} d_{\mathcal{H}}\left[\mathcal{D}_{T},\pi_{S} \mathcal{D}_{S}+\pi_{U} \mathcal{D}_{U}\right].
\end{equation}
Then, the generalization upper-bound for the target error $\epsilon_{T}(h)$ could be derived using the previous domain adaptation methods\cite{albuquerque2019generalizing,ben2006analysis,zhao2018adversarial}.

\newtheorem{theorem}{\bf Theorem}
\begin{theorem}
\label{thm1}
\emph{For a given classification hypothesis $h$ ($\forall h \in \mathcal{H}$), the target error $\epsilon_{T}(h)$ is bounded as}
\begin{equation}
\label{Eq:targetbound_triple}
\begin{split}
\epsilon_{T}(h) &\leq \epsilon_{S}(h)+\min \left\{\mathbb{E}_{\overline{\mathcal{D}}_{T}}\left[\left|F_{\overline{T}}-F_{T}\right|\right], \mathbb{E}_{\mathcal{D}_{T}}\left[\left|F_{T}-F_{\overline{T}}\right|\right]\right\}\\&+\pi_{U}(d_{\mathcal{H}}\left(\mathcal{D}_{S}, \mathcal{D}_{U}\right)+d_{\mathcal{H}}\left(\mathcal{D}_{U}, \mathcal{D}_{T}\right))+\pi_{S}d_{\mathcal{H}}\left(\mathcal{D}_{S}, \mathcal{D}_{T}\right)\\&+\pi_U\min \left\{\mathbb{E}_{\mathcal{D}_{S}}\left[\left|F_{S}-F_{U}\right|\right], \mathbb{E}_{\mathcal{D}_{U}}\left[\left|F_{U}-F_{S}\right|\right]\right\},
\end{split}
\end{equation}
\emph{where $\epsilon_{S}(h)$  is the error of the labeled source domain and $d_{\mathcal{H}}(\cdot)$ represents the $\mathcal{H}$-divergence between the given domains. $F_{\overline{T}}(x)= \pi_{S} F_{S}(x)+\pi_{U} F_{U}(x)$ is the labeling function for any $x \in Supp(\overline{D}_T)$. $F_{T}(x)$ is the labeling function of the target domain. Note that the labeling function in the present study includes the feature extractor $f(\cdot)$ and emotion classifier $Cls(\cdot)$. }
\end{theorem}
\emph{Proof} see Supplementary Materials Appendix B.

Under an extreme case where all the samples in the source domain are labeled ($\mathbb{U}$ is empty), the bound in Theorem 1 is equal to the bound given in Eq. \ref{Eq:targetbound}. Next, we will show how to optimize the target error empirically in the proposed semi-supervised multi-domain adaptation method. For an ideal joint hypothesis across $\mathbb{S}$, $\mathbb{U}$, and $\mathbb{T}$ domains, the second term ($\min \left\{\mathbb{E}_{\overline{\mathcal{D}}_{T}}\left[\left|F_{\overline{T}}-F_{T}\right|\right], \mathbb{E}_{\mathcal{D}_{T}}\left[\left|F_{T}-F_{\overline{T}}\right|\right]\right\}$) and the last term ($\pi_U\min \left\{\mathbb{E}_{\mathcal{D}_{S}}\left[\left|F_{S}-F_{U}\right|\right], \mathbb{E}_{\mathcal{D}_{U}}\left[\left|F_{U}-F_{S}\right|\right]\right\}$) in Theorem 1 are assumed to be small under the covariate shift assumption\cite{david2010impossibility,ben2010theory,albuquerque2019generalizing,sicilia2021domain}. 

\textbf{Empirical minimization of the $\mathcal{H}$-divergence.} Based on Theorem 1, it is found that a minimization of the $\mathcal{H}$-divergence among the domains $\mathbb{S}$, $\mathbb{U}$ and $\mathbb{T}$ is critical for the optimization of the target error $\epsilon_{T}(h)$. Previous studies have proved that the empirical $\mathcal{H}$-divergence can be approximated by the classification error of the domain discriminator\cite{ganin2016domain,ben2006analysis}. In other words, the minimization of the $\mathcal{H}$-divergence can be achieved by maximizing the domain discriminator loss via adversarial training. Considering the distribution shift among the given three domains, we re-define the domain discriminator loss as
\begin{equation}
\label{Eq:Tripledomainloss}
\begin{aligned}
\mathcal{L}_{\text{disc}}\left(\theta_{f}, \theta_{d}\right)=-\sum_{x} p\left(x\right) \log d\left(f\left(x\right)\right),
\end{aligned}
\end{equation}
where $p\left(x\right)$ is a one-hot domain label, referring to the belonging domain of the given EEG sample $x$. $d(\cdot)$ is a 3-class domain discriminator with parameter $\theta_d$, which outputs the domain prediction $d(f(x))$.

\textbf{Empirical minimization of the source error $\epsilon_{S}(h)$.} Instead of using the pointwise learning loss (e.g. cross-entropy loss) in the existing studies, we introduce to minimize the instance-wise pairwise learning loss as defined in Eq. \ref{Eq:insancepairwiseloss}.

\textbf{Empirical minimization of the target error $\epsilon_{T}(h)$.} The final objective loss of the proposed EEGMatch is defined as
\textcolor{black}{\begin{equation}
\label{Eq:finalloss_semi}
\mathcal{L}=\min _{\theta_{f},B}\max _{\theta_{d}}\mathcal{L}_{\text{pair}}(\theta_f,B)-\lambda\mathcal{L}_{\text{disc}}\left(\theta_{f}, \theta_{d}\right),
\end{equation}}
where $\mathcal{L}_{\text{pair}}(\theta_f,B)$ is the instance-wise pairwise learning loss given in Eq. \ref{Eq:insancepairwiseloss}, and $\mathcal{L}_{\text{disc}}\left(\theta_{f}, \theta_{d}\right)$ is the domain discriminator loss defined in Eq. \ref{Eq:Tripledomainloss}. $\lambda$ is a balanced hyperparameter to ensure the stability of adversarial training\cite{ganin2016domain}, defined as
\begin{equation}
    \lambda=\frac{2}{1+\exp (-\xi)}-1,
\end{equation}
where $\xi$ is a factor related to the training epoch, which is equal to the ratio of the current epoch to the maximum epoch. 

The overview of the training details of the EEGMatch is shown in the Supplementary Materials Algorithm S1. During the training process of the feature extractor $f \left(\cdot\right)$ and the domain discriminator $d \left(\cdot\right)$, the model is optimized with a gradient reversal layer (GRL)\cite{ganin2016domain}. The GRL layer can realize the adversarial training by reversing the gradient passed backward from $d \left(\cdot\right)$ to $f \left(\cdot\right)$. In other words, based on the empirical minimization of the $\mathcal{H}$-divergence (maximizing the domain discriminator loss $\mathcal{L}_{\text{disc}}\left(\theta_{f}, \theta_{d}\right)$) and the source error $\epsilon_{S}(h)$ (minimizing the pairwise loss $\mathcal{L}_{\text{pair}}(\theta_f,B)$), the target error $\epsilon_{T}(h)$ could be gradually minimized by decreasing the upper-bound in the training process as stated in Theorem 1.

\section{Experimental Results}
\label{sec:experiment}

\subsection{Benchmark databases}
To evaluate the effectiveness of the proposed EEGMatch, we conduct extensive experiments on three well-known public emotional EEG databases: SEED\cite{zheng2015investigating}, SEED-IV\cite{zheng2018emotionmeter} and SEED-V\cite{liu2021comparing}. In the SEED database, a total of three emotions (negative, neutral, and positive) were elicited using 15 movie clips, and the simultaneous EEG signals under the three emotional states were recorded from 15 subjects (7 males and 8 females) using the 62-channel ESI Neuroscan system. In the SEED-IV database, a total of four emotions (happiness, sadness, fear, and neutral) were elicited using 24 movie clips, and the simultaneous EEG signals under the four emotional states were also recorded from 15 subjects (7 males and 8 females) using the 62-channel ESI Neuroscan system. Each subject attended three separate sessions, and each session included 24 trials corresponding to 24 different movie clips. In the SEED-V database, a total of five emotions (happiness, sadness, fear, disgust, and neutral) were elicited using 15 movie clips, and the simultaneous EEG signals under the five emotions were recorded from 16 subjects (6 males and 10 females) using the 62-channel ESI Neuroscan system. Each subject attended three separate sessions, and each session included 15 trials corresponding to 15 different movie clips.

To make a fair comparison with the existing studies on the two benchmark databases, we also use the pre-computed differential entropy (DE) features as the model input. Specifically, for each trial, the EEG data was divided into a number of 1-second segments, and the DE features were extracted from each 1-second segment at the given five frequency bands (Delta, Theta, Alpha, Beta, and Gamma) from the 62 channels. Then, for each 1-second segment, it was represented by a 310-dimensional feature vector (5 frequency bands × 62 channels), which was further filtered by a linear dynamic system method for smooth purpose as the literature \cite{LDS2010,zheng2015investigating}.

\subsection{Implementation details and model setting}
 In the model implementation, the feature extractor $f(\cdot)$ and the domain discriminator $d(\cdot)$ are composed of fully-connected layers with the Relu activation function. Specifically, the feature extractor $f(\cdot)$ is designed as 310 neurons (input layer)-64 neurons (hidden layer 1)-Relu activation-64 neurons (hidden layer 2)-Relu activation-64 neurons (output feature layer). The domain discriminator $d(\cdot)$ is designed as 64 neurons (input layer)-64 neurons (hidden layer 1)-Relu activation-dropout layer-64 neurons (hidden layer 2)-3 neurons (output layer)-Softmax activation. The size of matrix $B$ given in Eq. \ref{Eq:bilinear} is $64 \times 64$. For the gradient descent and parameter optimization, we adopt the RMSprop optimizer. The learning rate is set to 1e-3 and the sample size of mini-batch is 48. All the trainable parameters are randomly initialized from a uniform distribution and no pretrained models are used. For the hyperparameters, we use $L2$ regularizes (1e-5) in the proposed model to avoid overfitting problems. In the EEG-Mixup based data augmentation, the coefficient $\alpha$ in the beta distribution is set to 0.5, and the ratio of the augmented sample size to the original sample size is set to 1:1. In the semi-supervised two-step pairwise learning, the temperature parameter $\eta$ used in Eq. \ref{Eq:sharpen} is set to 0.9 for SEED database and 0.8 for SEED-IV database. The dynamic upper and lower thresholds ($\tau_h^0$ and $\tau_l^0$) used in Eq. \ref{Eq:PairTargetNonLinear} are initialized as 0.9 and 0.5, respectively. The balance parameter $\gamma$ used in Eq. \ref{Eq:insancepairwiseloss} is controlled by a constant factor $\delta$ of 2. The regularization coefficient $\beta$ used in Eq. \ref{Eq:insancepairwiseloss} is set to 0.01. For all the compared deep learning methods reproduced in our experiments, we also use the same set of common hyperparameters and EEG differential entropy features as the proposed EEGMatch to ensure fair comparisons, such as the learning rate and the batch size. For the deep domain adaptation methods DANN\cite{ganin2016domain}, DCORAL\cite{Dcoral2016}, DDC\cite{DDC2014}, DAN\cite{He2018DAN}, we use the same feature extractor $f(\cdot)$ and domain discriminator $d(\cdot)$ as the proposed EEGMatch. The architecture of the classifier used in the methods above is designed as 64 neurons (hidden layer 1)- $C$ neurons (output layer), where $C$ is the number of classes. For the model architecture of the compared deep semi-supervised methods MixMatch\cite{berthelot2019mixmatch}, AdaMatch\cite{berthelot2021adamatch}, FlexMatch\cite{zhang2021flexmatch}, SoftMatch\cite{sohn2020fixmatch}, PARSE\cite{zhang2022parse}, we use the optimal settings introduced in \cite{zhang2022parse}. The model-specific hyperparameters used in different semi-supervised methods follow the optimal settings provided by the referenced literature. For the other compared methods, such as TCA\cite{TCA2010}, we follow the optimal setting and training pipelines introduced in the referenced literature.

\subsection{Experimental protocol with incomplete labels}
We also adopt the cross-subject leave-one-subject-out experimental protocol as the existing studies\cite{song2018eeg,BiHDM2019,zhong2020eeg,Du2020} for cross-comparison. Specifically, we treat 14 subjects as the source domain and the remaining 1 subject as the target domain. For the source domain, only the first $N$ trials of each subject in one session have labels (the labeled source domain $\mathbb{S}$), and the rest $N_{max}-N$ trials of each subject are without labels (the unlabeled source domain $\mathbb{U}$). The remaining 1 subject's $N_{T}$ trials are considered as the target domain ($\mathbb{T}$) for transfer learning and model evaluation. In the domains $\mathbb{S}$, $\mathbb{U}$ and $\mathbb{T}$, the number of original trials used for EEG-Mixup augmentation is $N$, $N_{max}-N$ and $N_{T}$, respectively. Given that the augmented sample ratio is set to 1:1, the total number of trials used for model training is $2\times(N+N_{max}-N+N_{T})=2\times(N_{max}+N_{T})$. Assuming that there are $Z$ samples within one original trial, the total number of samples for transfer learning is $2\times Z\times(N_{max}+N_{T}$). Note that only the original samples in the target domain $\mathbb{T}$ are used for model evaluation. We repeat $T$ times until each subject is treated as the target domain for once, where $T$ is the number of subjects in the database. Then we compute the averaged accuracy and the standard deviation as the final model performance. It is noted that the label information of $\mathbb{U}$ and $\mathbb{T}$ are unknown during model training. Furthermore, to measure the model stability under different incomplete label conditions, we adjust $N$ at different values. Here, considering the SEED and SEED-V databases have 15 trials and the SEED-IV database has 24 trials for each subject in one session, the $N$ value varies from 3 to 12, 5 to 10, and from 4 to 20 for the SEED, SEED-V, and SEED-IV databases, respectively. To ensure consistency with a state-of-the-art model in semi-supervised emotion recognition \cite{zhang2022parse}, we adopt a similar experimental setup, where the first session of the SEED database and three sessions from the SEED-IV and SEED-V databases are used for experiments.

\subsection{Cross-subject leave-one-subject-out cross-validation results with incomplete labels}
Table \ref{tab:seedsemicompare} reports the experimental results on the SEED database, in a comparison with the existing popular machine learning or deep learning methods. The results show that the proposed EEGMatch achieves superior performance compared to the other models under different incomplete label conditions, where the average model performance across $N=3, 6, 9, 12$ is 91.35$\pm$07.03. Compared to the performance achieved by the best competitor in the literature (DDC\cite{DDC2014}: 85.46$\pm$08.34), the average performance enhancement under different $N$ values is 5.89\%. It demonstrates that, even with only a few labeled data (when $N$ is small), EEGMatch still can well utilize both the labeled source data ($\mathbb{S}$) and the unlabeled source data ($\mathbb{U}$) to enhance the model generalization ability and reach a satisfying emotion recognition performance in the target domain ($\mathbb{T}$).

The experimental results on the SEED-IV and SEED-V databases under different incomplete label conditions are reported in Table \ref{tab:seedivsemicompare} and Table \ref{tab:seedvsemicompare}, which also show the superiority of the proposed EEGMatch on the semi-supervised cross-subject emotion recognition task. Specifically, the average model performance on the SEED-IV database across $N=4, 8, 12, 16, 20$ is 65.53$\pm$08.31. In addition, the average model performance on the SEED-V database across $N=5, 15$ is 62.75$\pm$09.02. Compared to the best results in the literature, the average performance enhancement under different $N$ values is 0.93\% for the SEED-IV database (DDC\cite{DDC2014}: 64.60$\pm$08.82) and 0.28\% for the SEED-V database (DDC\cite{DDC2014}: 62.47$\pm$08.25). On the other hand, the improvement of the experimental performance on the two databases is less obvious than that on the SEED database. One possible reason is that more emotion classes in the databases would make the predicted pseudo labels ($\hat{Y}_{u}$) more unstable.

\begin{table*}[]
\begin{center}
\caption{The mean accuracies (\%) and standard deviations (\%) of emotion recognition results using cross-subject leave-one-subject-out cross-validation on the SEED database with incomplete labels. For each subject, $N$ trials have labels and $15-N$ trials do not have labels. The model results reproduced by us are indicated by `*'.}
\label{tab:seedsemicompare}
\scalebox{1}{
\begin{tabular*}{\hsize}{@{}@{\extracolsep{\fill}}lccccc@{}}
\toprule
Methods   & $N=3$    & $N=6$   & $N=9$ & $N=12$ & Average  \\ 
\midrule
SVM*\cite{SVM1999}         & 65.26$\pm$09.67     & 70.78$\pm$08.31     &70.61$\pm$08.72  &68.91$\pm$10.56 &68.89$\pm$09.61\\
TCA*\cite{TCA2010}         & 58.68$\pm$09.66     & 60.20$\pm$09.63     &59.93$\pm$10.63  &60.20$\pm$10.59 &59.75$\pm$10.16\\
SA*\cite{SA2013}           & 48.79$\pm$10.69     & 49.68$\pm$14.16     &54.69$\pm$10.35  &55.94$\pm$08.29 &52.27$\pm$11.50\\
KPCA*\cite{KPCA1999}       & 57.92$\pm$10.36     & 58.06$\pm$08.73     &59.39$\pm$09.79  &60.73$\pm$09.36 &59.03$\pm$09.65\\
RF*\cite{Breiman2001RF}    & 59.58$\pm$11.24     & 63.14$\pm$09.63     &67.77$\pm$07.83  &70.89$\pm$07.98 &65.35$\pm$10.23\\
Adaboost*\cite{2006Boost}  & 62.45$\pm$09.94     & 68.03$\pm$09.78     &71.72$\pm$08.36  &72.59$\pm$09.52 &68.70$\pm$10.23\\
CORAL*\cite{CORAL2016}     & 66.21$\pm$09.85     & 70.16$\pm$08.70     &73.28$\pm$10.30  &73.97$\pm$10.25 &70.90$\pm$10.27\\
GFK*\cite{GFK2012}         & 58.11$\pm$11.56     & 60.12$\pm$10.70     &60.85$\pm$08.94  &60.55$\pm$10.02 &59.91$\pm$10.40\\
KNN*\cite{KNN1982}         & 57.47$\pm$11.40     & 59.87$\pm$10.08     &60.71$\pm$08.75  &61.38$\pm$09.56 &59.86$\pm$10.10\\
DGCNN*\cite{song2018eeg}& 71.75$\pm$09.73 & 79.26$\pm$07.00 & 78.21$\pm$07.27 & 79.11$\pm$07.79 & 77.08$\pm$08.60 \\	
DAN*\cite{He2018DAN}      & 78.57$\pm$14.40     & 84.32$\pm$10.30     &85.89$\pm$08.29  &86.31$\pm$08.24 &83.77$\pm$11.05\\
DANN*\cite{ganin2016domain}        & 77.41$\pm$09.84     & 83.93$\pm$09.42     &85.75$\pm$07.21  &86.74$\pm$07.09 &83.46$\pm$09.23\\
DCORAL*\cite{Dcoral2016}   & 74.42$\pm$16.02     & 82.01$\pm$09.03     &84.00$\pm$06.39  &83.15$\pm$06.98 &80.89$\pm$11.02\\
DDC*\cite{DDC2014}         & 80.82$\pm$09.37     & 85.40$\pm$09.27     &88.06$\pm$04.75  &87.58$\pm$07.00 &85.46$\pm$08.34\\
PARSE*\cite{zhang2022parse}& 76.34$\pm$07.58 & 79.46$\pm$04.63 &81.21$\pm$04.19 &81.73$\pm$04.99 &79.68$\pm$05.90 \\		
MixMatch*\cite{berthelot2019mixmatch} & 63.66$\pm$09.85 & 77.62$\pm$05.48 &83.31$\pm$04.40 &81.12$\pm$05.17 &76.43$\pm$10.09 \\
AdaMatch*\cite{berthelot2021adamatch} & 74.83$\pm$05.71 & 78.67$\pm$08.52 &80.90$\pm$07.31 &81.69$\pm$07.03 &79.02$\pm$07.69 \\
FlexMatch*\cite{zhang2021flexmatch} & 76.22$\pm$09.09 & 80.65$\pm$06.70 & 84.04$\pm$05.38 & 84.35$\pm$06.41 & 81.32$\pm$07.76\\
SoftMatch*\cite{chen2023softmatch} & 74.99$\pm$09.33 & 80.09$\pm$06.67 & 83.84$\pm$04.86 & 83.98$\pm$05.66 & 80.72$\pm$07.76 \\
\midrule
\textbf{EEGMatch} & \textbf{86.68$\pm$07.89} & \textbf{92.70$\pm$06.37}     & \textbf{93.61$\pm$04.03}     &\textbf{92.39$\pm$06.97}  &\textbf{91.35$\pm$07.03}\\
\bottomrule
\end{tabular*}}
\end{center}
\end{table*}

\begin{table*}[]
\begin{center}
\caption{The mean accuracies (\%) and standard deviations (\%) of emotion recognition results using cross-subject leave-one-subject-out cross-validation on the SEED-IV database with incomplete labels. For each subject, $N$ trials have labels and $24-N$ trials do not have labels. The model results reproduced by us are indicated by `*'.}
\label{tab:seedivsemicompare}
\scalebox{1}{
\begin{tabular*}{\hsize}{@{}@{\extracolsep{\fill}}lcccccc@{}}
\toprule
Methods   & $N=4$    & $N=8$   & $N=12$ & $N=16$ & $N=20$ & Average   \\ 
\midrule
SVM*\cite{SVM1999}         & 45.30$\pm$12.26 & 50.40$\pm$14.54 &52.59$\pm$10.30 &47.75$\pm$11.34 &49.76$\pm$12.10 &49.16$\pm$12.44\\
TCA*\cite{TCA2010}         & 32.57$\pm$07.25 & 40.63$\pm$10.34 &42.37$\pm$11.42 &42.59$\pm$10.50 &41.82$\pm$11.03 &40.00$\pm$10.89\\
SA*\cite{SA2013}           & 29.86$\pm$04.90 & 41.62$\pm$12.38 &39.45$\pm$07.12 &35.88$\pm$13.08 &39.72$\pm$10.62 &37.31$\pm$10.94\\
KPCA*\cite{KPCA1999}       & 35.02$\pm$06.92 & 40.84$\pm$09.90 &42.18$\pm$10.15 &43.77$\pm$10.24 &42.44$\pm$12.33 &40.85$\pm$10.51\\
RF*\cite{Breiman2001RF}    & 41.51$\pm$10.99 & 45.36$\pm$12.58 &49.73$\pm$09.64 &51.42$\pm$14.19 &49.70$\pm$14.56 &47.54$\pm$13.05\\
Adaboost*\cite{2006Boost}  & 41.37$\pm$09.57 & 52.37$\pm$13.39 &52.30$\pm$13.51 &52.15$\pm$13.96 &51.96$\pm$15.69 &50.03$\pm$14.06\\
CORAL*\cite{CORAL2016}     & 43.64$\pm$11.68 & 50.62$\pm$13.88 &52.66$\pm$13.97 &49.89$\pm$13.42 &51.17$\pm$12.60 &49.60$\pm$13.50\\
GFK*\cite{GFK2012}         & 35.30$\pm$07.76 & 44.46$\pm$13.03 &47.01$\pm$12.12 &44.44$\pm$10.10 &41.94$\pm$09.73 &42.63$\pm$11.43\\
KNN*\cite{KNN1982}         & 35.88$\pm$07.67 & 42.75$\pm$11.19 &43.51$\pm$10.80 &41.74$\pm$07.48 &37.92$\pm$09.06 &40.36$\pm$09.82\\
DGCNN*\cite{song2018eeg}& 53.69$\pm$04.70 & 60.25$\pm$07.68 & 63.18$\pm$09.19 & 64.07$\pm$08.15 & 62.88$\pm$09.55 & 60.81$\pm$08.89\\
DAN*\cite{He2018DAN}      & 55.84$\pm$07.37 & 62.78$\pm$08.55 &65.59$\pm$10.71 &64.61$\pm$09.04 &66.19$\pm$07.69 &63.00$\pm$09.53\\
DANN*\cite{ganin2016domain}        & 55.36$\pm$08.11 & 63.74$\pm$10.89 &66.32$\pm$09.77 &65.55$\pm$09.94 &66.20$\pm$09.19 &63.43$\pm$10.48\\
DCORAL*\cite{Dcoral2016}   & 56.56$\pm$07.97 & 62.49$\pm$09.42 &65.65$\pm$09.69 &64.34$\pm$08.38 &65.03$\pm$08.84 &62.81$\pm$09.48\\
DDC*\cite{DDC2014}         & 56.80$\pm$07.91 & 63.88$\pm$08.55 &\textbf{67.94$\pm$06.60} &67.34$\pm$07.97 &67.03$\pm$07.77 &64.60$\pm$08.82\\
PARSE*\cite{zhang2022parse}& 59.12$\pm$07.74 &62.34$\pm$08.54 &65.24$\pm$08.88 &66.47$\pm$08.57 &66.82$\pm$07.91 &64.00$\pm$08.83\\
MixMatch*\cite{berthelot2019mixmatch} & 57.51$\pm$06.16 & 62.12$\pm$05.82 &64.11$\pm$07.92 &65.34$\pm$08.39 &65.95$\pm$08.42 &63.01$\pm$08.03\\
AdaMatch*\cite{berthelot2021adamatch} & 57.90$\pm$08.55 & 62.67$\pm$08.25 &64.50$\pm$09.35 &66.17$\pm$08.31 &\textbf{67.44$\pm$08.65} &63.74$\pm$09.25\\
FlexMatch*\cite{zhang2021flexmatch} & 57.59$\pm$07.73 & 60.69$\pm$08.50 & 62.10$\pm$09.55 & 63.24$\pm$09.74 & 63.67$\pm$08.93 & 61.46$\pm$09.19\\
SoftMatch*\cite{chen2023softmatch} & 58.18$\pm$07.70 & 61.16$\pm$08.81 & 62.15$\pm$09.29 & 62.95$\pm$09.39 & 63.50$\pm$08.96 & 61.59$\pm$09.05\\
\midrule
\textbf{EEGMatch} & \textbf{60.58$\pm$07.38} & \textbf{65.66$\pm$08.75}     & 67.40$\pm$09.30     &\textbf{67.75$\pm$07.16}  &66.29$\pm$06.55  & \textbf{65.53$\pm$08.31} \\
\bottomrule
\end{tabular*}}
\end{center}
\end{table*}

\begin{table*}[]
\begin{center}
\caption{The mean accuracies (\%) and standard deviations (\%) of emotion recognition results using cross-subject leave-one-subject-out cross-validation on the SEED-V database with incomplete labels. For each subject, $N$ trials have labels and $15-N$ trials do not have labels. The model results reproduced by us are indicated by `*'.}
\label{tab:seedvsemicompare}
\scalebox{1}{
\begin{tabular*}{\hsize}{@{}@{\extracolsep{\fill}}lccclccc@{}}
\toprule
Methods   & $N=5$    & $N=10$   & Average & Methods   & $N=5$    & $N=10$   & Average\\ 
\midrule
SVM*\cite{SVM1999}  &45.46$\pm$10.19	&45.46$\pm$10.85	&45.46$\pm$10.69
&TCA*\cite{TCA2010} &30.73$\pm$07.96	&33.17$\pm$08.41	&31.95$\pm$08.41\\
SA*\cite{SA2013}    &25.90$\pm$01.13	&30.19$\pm$06.06	&28.05$\pm$04.94 &KPCA*\cite{KPCA1999} &29.13$\pm$08.64	&32.91$\pm$11.44	&31.02$\pm$10.48\\
RF*\cite{Breiman2001RF} &36.42$\pm$10.11	&39.42$\pm$07.58	&37.92$\pm$09.21
&Adaboost*\cite{2006Boost} &40.00$\pm$11.82	&41.95$\pm$09.68	&40.98$\pm$11.02\\
CORAL*\cite{CORAL2016} &44.73$\pm$09.89	&45.14$\pm$10.67	&44.94$\pm$10.45
&GFK*\cite{GFK2012}  &31.37$\pm$07.86	&35.63$\pm$08.92	&33.50$\pm$08.81\\
KNN*\cite{KNN1982}  &31.61$\pm$08.44	&34.64$\pm$08.70	&33.13$\pm$08.84
&DAN*\cite{He2018DAN} &54.56$\pm$07.05	&63.95$\pm$09.04	&59.25$\pm$09.37\\
DANN*\cite{ganin2016domain}  &55.04$\pm$07.74	&64.01$\pm$09.77	&59.53$\pm$09.89
&DCORAL*\cite{Dcoral2016} &56.65$\pm$08.83	&62.90$\pm$08.00	&59.78$\pm$08.98\\
DDC*\cite{DDC2014} &58.67$\pm$07.23	&66.26$\pm$07.41	&62.47$\pm$08.25
&DGCNN*\cite{song2018eeg} &	45.49$\pm$07.17&	54.38$\pm$11.79 & 49.93$\pm$10.72\\
MixMatch*\cite{berthelot2019mixmatch} &52.83$\pm$06.07	&63.32$\pm$08.93	&58.08$\pm$09.26
&AdaMatch*\cite{berthelot2021adamatch} &58.25$\pm$05.19	&65.71$\pm$07.41	&61.98$\pm$07.41\\
FlexMatch*\cite{zhang2021flexmatch} &	57.48$\pm$11.71 & 65.06$\pm$13.38	& 61.27$\pm$13.13
&SoftMatch*\cite{chen2023softmatch}  &57.74$\pm$11.91	&64.81$\pm$12.55	&61.27$\pm$12.74\\
PARSE*\cite{zhang2022parse}  &56.00$\pm$08.31	&65.50$\pm$08.18	&60.75$\pm$09.52
&\textbf{EEGMatch}&	\textbf{59.06$\pm$07.45} &	\textbf{66.44$\pm$08.94} & \textbf{62.75$\pm$09.02}\\
\bottomrule
\end{tabular*}}
\end{center}
\end{table*}

\section{Discussion and Conclusion}
\label{sec:discussion}
To fully analyze the performance and the robustness of the proposed model, we conduct various experiments to evaluate the contribution
of each module in the EEGMatch and discuss the effect of different hyperparameter settings. All the present results in this session are based on the SEED database.

\subsection{Ablation Study}
In this section, we conduct the ablation study to investigate the contributions of different modules in the proposed model. Table \ref{tab:seedsemiablation} reports the cross-subject leave-one-subject-out emotion recognition results on the SEED database under different incomplete label conditions. (1) When the EEG-Mixup based data augmentation module is removed from the framework, the recognition performance presents a significant decrease for all incomplete label conditions. The average model performance drops from 91.35$\pm$07.03 to 89.23$\pm$07.76. It shows that the EEG-Mixup based data augmentation can benefit the model performance by generating more valid data for model training. We also replace the proposed EEG-Mixup based data augmentation with the standard Mixup based data augmentation\cite{zhang2018mixup} and find that the standard Mixup based data augmentation has a negative impact on the performance due to the ignorance of the IID assumption. The average model performance under different $N$ values drops by 3.57\%. Note that the number of original trials used for Mixup augmentation is the same as the proposed EEG-Mixup. The augmented sample ratio is also set to 1:1 for both EEG-Mixup and Mixup augmentation methods. (2) Comparing the model performance with and without the semi-supervised prototype-wise pairwise learning, an obvious drop could be observed (from 91.35$\pm$07.03 to 88.85$\pm$08.89) when the semi-supervised prototype-wise pairwise learning is removed from the proposed model framework. The benefit of prototypical learning has also been observed in the other few-shot learning or transfer learning studies\cite{Pinheiro2018,Prototype2017,MixPrototypes2019,zhou2023pr}. (3) The significant positive impact of the introduced semi-supervised instance-wise pairwise learning on the model performance is observed, with an average improvement in emotion recognition performance (7.20\%) being noted under various $N$ values. The results suggest that the semi-supervised instance-wise pairwise learning can help the model capture inherent structures among EEG samples and extract representative and informative features for the downstream classification tasks. (4) To evaluate the validity of the proposed semi-supervised multi-domain adaptation method, we replace it using the traditional DANN-based domain adaptation ($\mathbb{U}$ is not considered in the domain adaptation process) and calculate the model performance under different incomplete label conditions. The results show that, compared to the traditional DANN-based domain adaptation, the proposed semi-supervised multi-domain adaptation method (EEGMatch) could estimate a more accurate upper-bound for the target error by considering the distribution shift over all multiple domains (labeled source domain $\mathbb{S}$, unlabeled source domain $\mathbb{U}$, and target domain $\mathbb{T}$). On the average, the proposed semi-supervised multi-domain adaptation method could enhance the model performance by 1.27\%. (5) Comparing the model performance with and without the progressive learning strategy, a remarkable performance improvement is observed when the strategy is applied. Specifically, the average model performance under different $N$ values increases by 1.45\%. Note that we fix the initial upper and lower thresholds throughout the model training process when the updating strategy is removed ($\tau_{h}^{0}=0.9$, $\tau_{l}^{0}=0.5$).  Compared with the fixed thresholds, dynamic and progressive thresholds can allow more samples to participate in the training process, which benefits the model generalization performance\cite{zhang2021flexmatch,zhou2023pr,chen2023softmatch}. 

\begin{table*}[]
\begin{center}
\caption{The ablation study of the proposed EEGMatch.}
\label{tab:seedsemiablation}
\scalebox{1}{
\begin{tabular*}{\hsize}{@{}@{\extracolsep{\fill}}lccccc@{}}
\toprule
Methods   & $N=3$    & $N=6$   & $N=9$ & $N=12$ & Average   \\ 
\midrule
Without EEG-Mixup based data augmentation            & 84.93$\pm$08.56     & 89.40$\pm$07.81     &89.59$\pm$06.86  &\textbf{93.01$\pm$05.17} &89.23$\pm$07.76\\
With standard Mixup based data augmentation & 84.93$\pm$08.05 & 88.25$\pm$09.38 & 89.32$\pm$07.69 &88.59$\pm$07.90 &87.78$\pm$08.45\\
Without semi-supervised prototype-wise pairwise learning  & 81.89$\pm$11.79     & 91.45$\pm$05.81     &90.56$\pm$04.53  &91.49$\pm$07.60 &88.85$\pm$08.89\\
Without semi-supervised instance-wise pairwise learning   & 81.04$\pm$07.47     & 83.09$\pm$08.40     &85.28$\pm$08.16  &87.19$\pm$08.57 &84.15$\pm$08.48\\
With traditional DANN-based domain adaptation & 82.69$\pm$10.72     & \textbf{93.29$\pm$05.54}     &92.87$\pm$06.73 &91.48$\pm$05.24 &90.08$\pm$08.56\\	
Without the progressive learning strategy  &83.19$\pm$10.79 &92.01$\pm$07.66 &92.67$\pm$05.50 &91.74$\pm$06.16 &89.90$\pm$08.71\\
\midrule
\textbf{EEGMatch} & \textbf{86.68$\pm$07.89} & 92.70$\pm$06.37     &\textbf{93.61$\pm$04.03}     &92.39$\pm$06.97  &\textbf{91.35$\pm$07.03}\\
\bottomrule
\end{tabular*}}
\end{center}
\end{table*}

\subsection{The training performance on different domains}
\begin{figure*}[h]
\begin{center}
\subfloat[]{\includegraphics[width=0.33\linewidth]{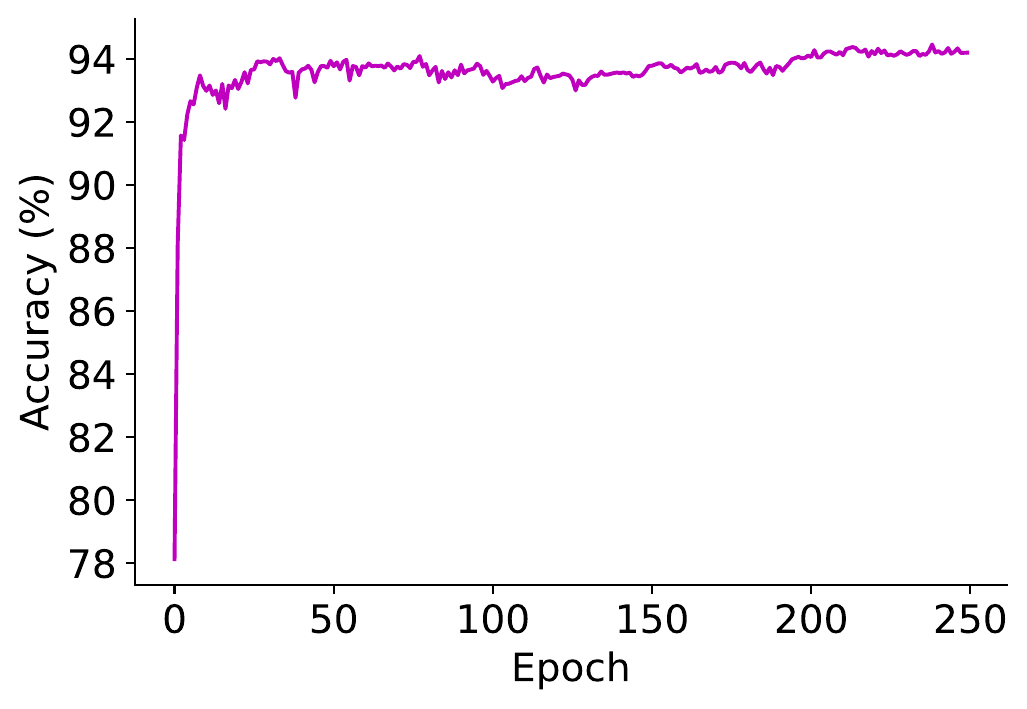}}
\subfloat[]{\includegraphics[width=0.33\linewidth]{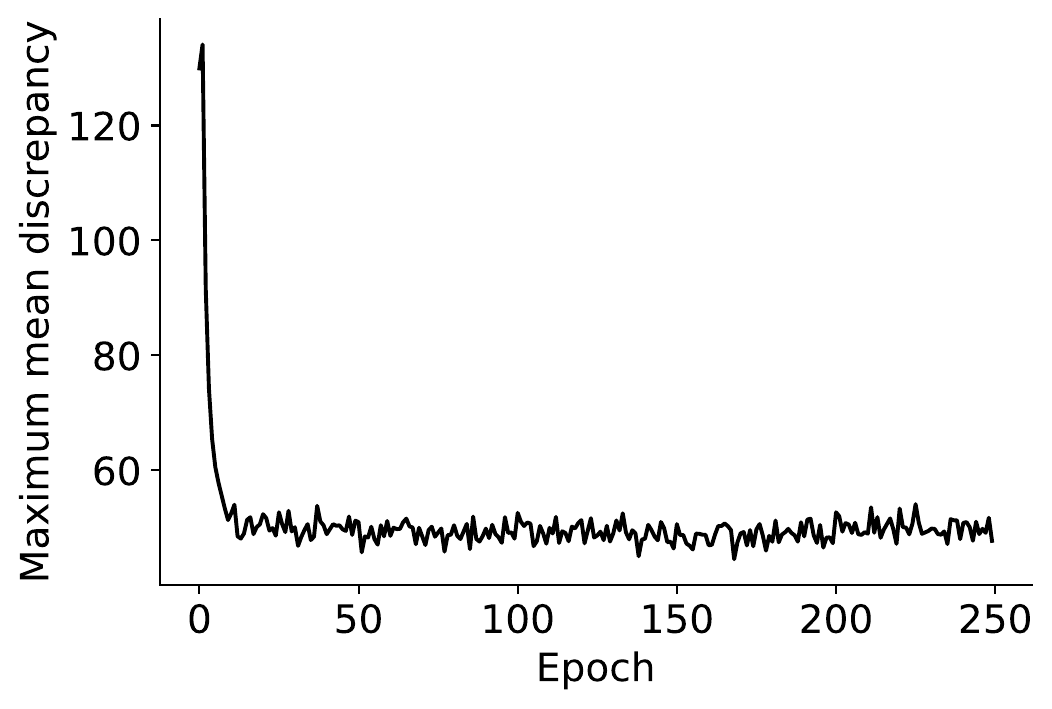}}
\subfloat[]{\includegraphics[width=0.33\linewidth]{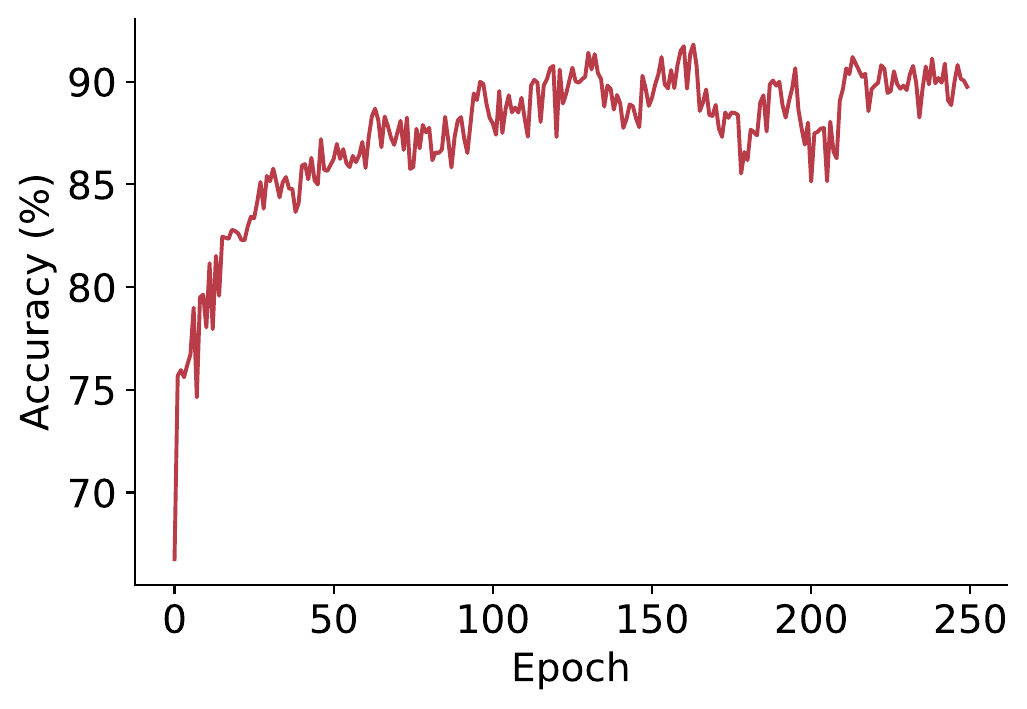}}
\caption{The training performance on different domains. (a) The training process on the source domain ($\mathbb{S}$+$\mathbb{U}$). (b) The estimated $\mathcal{H}$-divergence among different domains in terms of the maximum mean discrepancy (MMD) calculation at different training epochs. (c) The training process on the target domain ($\mathbb{T}$).}
\label{fig:curve}
\end{center}
\end{figure*}

The training process of the proposed EEGMatch on different domains is reported in Fig. \ref{fig:curve} ($N=12$). Here, we present the average training performance across different subjects to show the source and target error given in Theorem \ref{thm1}. A higher classification accuracy refers to a lower modeling error. To estimate the $\mathcal{H}$-divergence among different domains ($\pi_{U}(d_{\mathcal{H}}\left(\mathcal{D}_{S}, \mathcal{D}_{U}\right)+d_{\mathcal{H}}\left(\mathcal{D}_{U}, \mathcal{D}_{T}\right))+\pi_{S}d_{\mathcal{H}}\left(\mathcal{D}_{S}, \mathcal{D}_{T}\right)$), we adopt the maximum mean discrepancy (MMD) \cite{sejdinovic2013equivalence} calculation here. The findings reveal that the target error $\epsilon_{T}(h)$ shown in Fig. \ref{fig:curve} (c) reduces along with a decline in the source error $\epsilon_{S}(h)$ shown in Fig. \ref{fig:curve} (a) and the $\mathcal{H}$-divergence among different domains shown in Fig. \ref{fig:curve} (b). We also observe that the average accuracy curve on the target domain exhibits some irregularities during model training. This could be influenced by the non-stationarity of the domain adversarial training\cite{arjovsky2017wasserstein}. A straightforward way to alleviate this problem is to adjust the learning rate. We carefully examine the impact of varying learning rates on mitigating the observed irregularities and report the investigation results in Supplementary Materials: Appendix G.

\begin{figure*}[h]
\begin{center}
\subfloat[]{\includegraphics[width=0.33\linewidth]{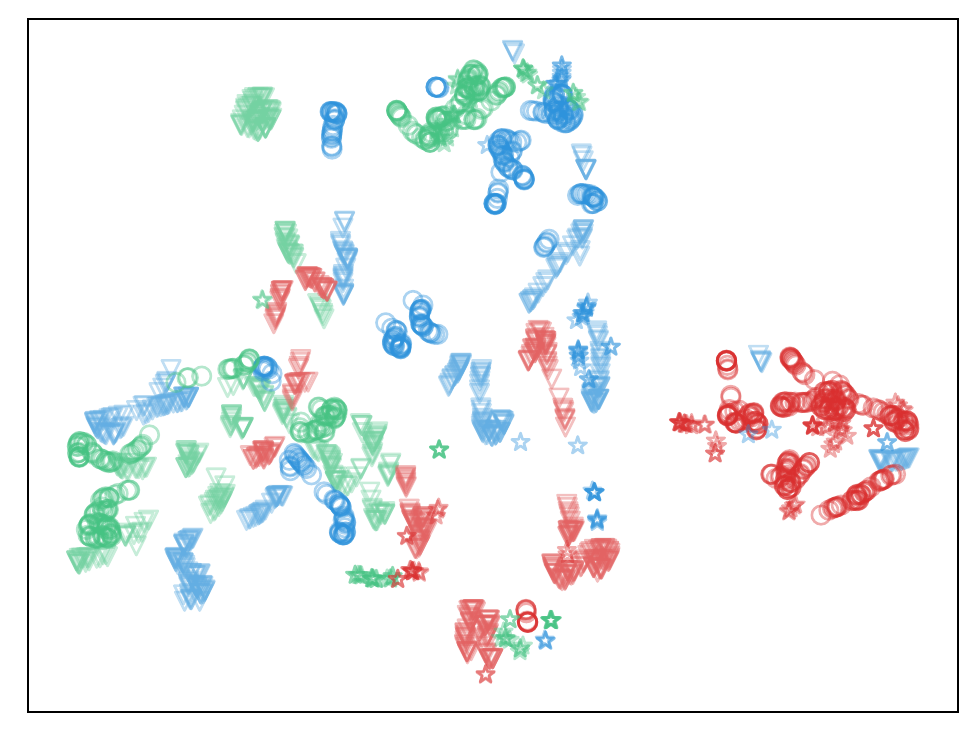}}
\subfloat[]{\includegraphics[width=0.33\linewidth]{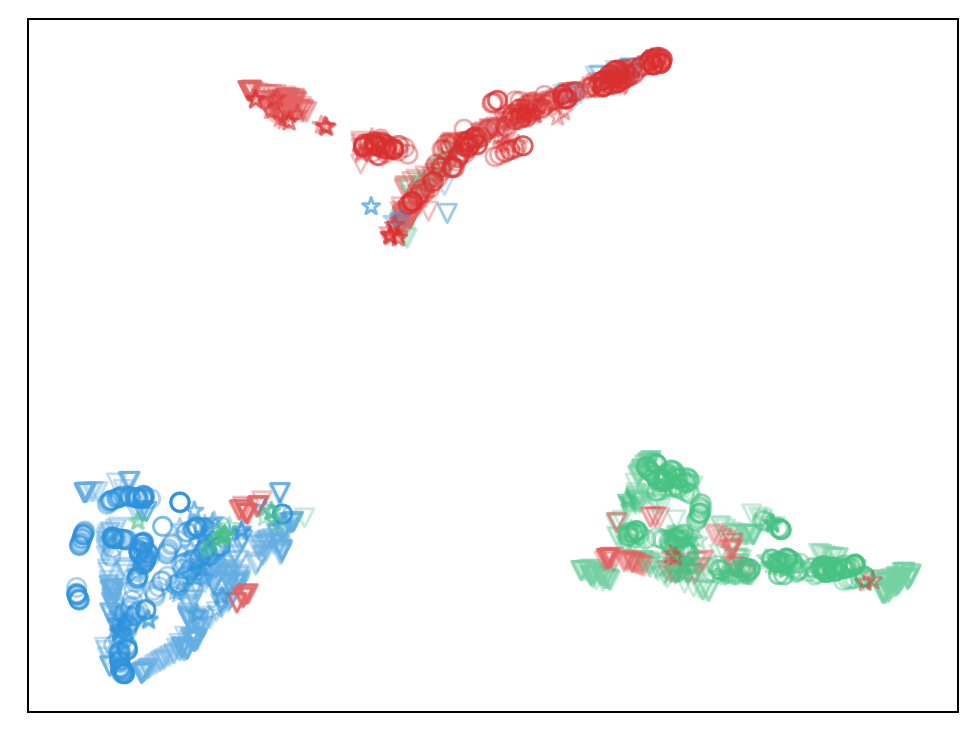}}
\subfloat[]{\includegraphics[width=0.33\linewidth]{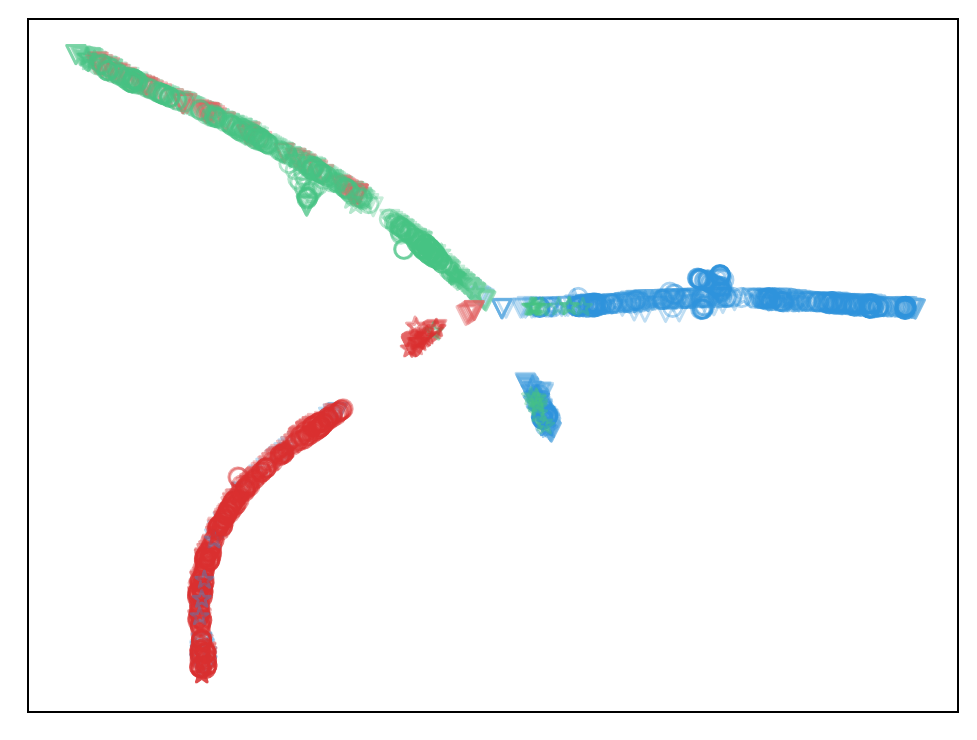}}
\caption{A visualization of the learned feature representations (a) before training, (b) at the training epoch of 50, and (c) in the final model. Here, the circle, asterisk and triangle represent the labeled source domain ($\mathbb{S}$), the unlabeled source domain ($\mathbb{U}$), and the target domain ($\mathbb{T}$). The blue, green, and red colors indicate negative, neutral, and positive emotions.}
\label{fig:tsne}
\end{center}
\end{figure*}

The learned feature representations by the feature extractor $f(\cdot)$ at different training stages are also visually compared, based on the t-distributed stochastic neighbor embedding (t-SNE) algorithm\cite{2008Tsne}. As shown in Fig. \ref{fig:tsne} (a), it is found that the distribution mismatch of the extracted features from different domains is quite obvious before model training. Along with the minimization of the source error and the $\mathcal{H}$-divergence defined in Theorem \ref{thm1}, the target error ($\epsilon_{T}(h)$) could be gradually minimized, as shown in Fig. \ref{fig:tsne} (b) and (c). In addition, the distribution mismatch between different domains is reduced, demonstrating that EEGMatch can address the covariate shift problem caused by the non-stationarity property and the individual differences in EEG signals. Our findings support the effectiveness of the proposed semi-supervised multi-domain adaptation in EEG modeling under a semi-supervised learning framework.

Overall, these results empirically verify the statement in Theorem \ref{thm1} that the target error $\epsilon_{T}(h)$ could be gradually minimized by decreasing the upper-bound in the training process.

\begin{table*}[h]
\setlength{\tabcolsep}{0.45em}
\begin{center}
\caption{The mean accuracies (\%) and standard deviations (\%) of emotion recognition on unseen target data. The model results reproduced by us are indicated by `*'.}
\label{tab:incompleteTargetData}
\scalebox{1}{
\begin{tabular*}{\hsize}{@{}@{\extracolsep{\fill}}llccccc@{}}
\toprule
Validation Set & Test Set & DAN*\cite{He2018DAN} & DANN*\cite{ganin2016domain} & DCORAL*\cite{Dcoral2016} & DDC*\cite{DDC2014} & \textbf{EEGMatch}\\ 
\midrule
$\mathbb{T}_{v} = 3$  trials &$\mathbb{T}_{t}$ = 12 trials  
& 81.92$\pm$06.35 & 79.50$\pm$06.46  & 81.02$\pm$06.36 & \textbf{82.99$\pm$06.98}
 & 81.97$\pm$07.22\\
$\mathbb{T}_{v} = 6$  trials & $\mathbb{T}_{t}$ = 9  trials
& 81.76$\pm$06.25 & 81.69$\pm$08.42 & 81.16$\pm$08.46 & 82.37$\pm$08.81
 &\textbf{84.73$\pm$06.25}\\
$\mathbb{T}_{v} = 9$  trials &$\mathbb{T}_{t}$ = 6 trials
& 82.66$\pm$06.26 & 81.26$\pm$07.74 & 80.48$\pm$07.70 & 82.54$\pm$06.85
 &\textbf{85.63$\pm$06.62}\\
$\mathbb{T}_{v} = 12$ trials &$\mathbb{T}_{t}$ = 3  trials 
& 84.40$\pm$14.23 & 80.26$\pm$14.52 & 82.23$\pm$07.91
& 84.65$\pm$12.47  &\textbf{85.72$\pm$10.30}\\
\midrule
\multicolumn{2}{l}{Average Performance} & 82.69$\pm$09.02 & 80.67$\pm$09.83 & 81.22$\pm$07.67 &  83.14$\pm$09.11 & \textbf{84.52$\pm$08.34} \\
\bottomrule
\end{tabular*}}
\end{center}
\end{table*}

\subsection{Model performance on unseen target data}
In the existing EEG-based emotion recognition models using transfer learning strategy, all the target data are assumed to be complete and available for domain adaptation\cite{Zheng2016,He2018DAN,JinDANNfisrt,Du2020,zhong2020eeg,zhang2022parse,he2022adversarial}. However, in real-world applications, modeling with all of the target data is impractical. To further evaluate the stability and reliability of the proposed EEGMatch on unseen data, we divide the target domain ($\mathbb{T}$) into a validation set ($\mathbb{T}_{v}$: the first $K$ trials in the target domain) and a test set ($\mathbb{T}_{t}$: the remaining $15-K$ in the target domain). Here, the validation set ($\mathbb{T}_{v}$) is used for domain adaptation, together with the labeled source data ($\mathbb{S}$) and the unlabeled source data ($\mathbb{U}$). Then, the trained model will be validated on the unseen test set ($\mathbb{T}_{t}$) to evaluate the model generalization capability. It should be noted that $\mathbb{U}$ contains the last three trials of the source subjects in this experiment ($N=12$). We compare the performance of the EEGMatch with its top-4 competitors in the SEED database and report the results with different $K$ values in Table \ref{tab:incompleteTargetData}. It can be observed that the proposed EEGMatch achieves the best average performance under different $K$ values, which demonstrates its effectiveness on the unseen data.

\subsection{The benefit of the introduced unlabeled source domain $\mathbb{U}$}
\begin{figure}[]
\begin{center}
\includegraphics[width=0.4\textwidth]{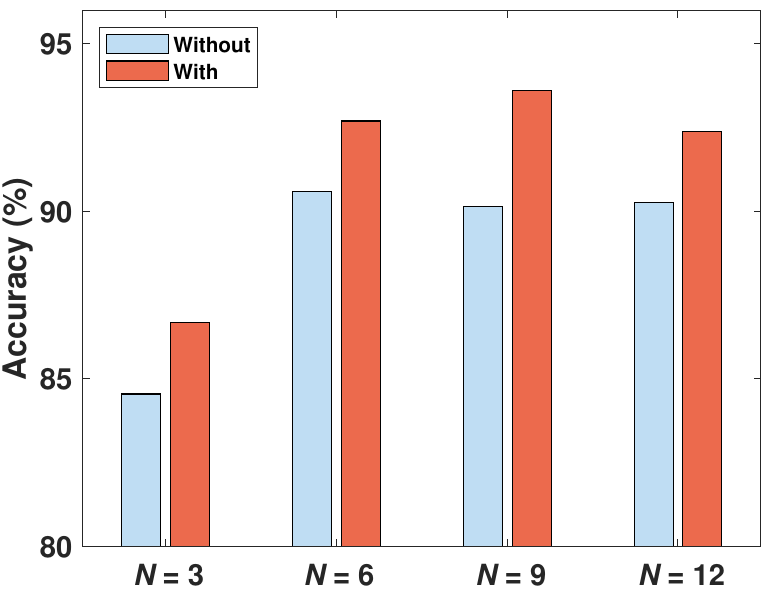}
\caption{The emotion recognition accuracies (\%) without and with the unlabeled source domain ($\mathbb{U}$) under different $N$ values.}
\label{fig:SEED_U}
\end{center}
\end{figure}

To further investigate the contribution of the introduced unlabeled source domain ($\mathbb{U}$) in solving the semi-supervised EEG-based emotion recognition task, we compare the model performance with and without $\mathbb{U}$ under different incomplete label conditions. Note that for the model without $\mathbb{U}$, the data augmentation, two-step pairwise learning, and multi-domain adaptation of $\mathbb{U}$ domain are not included. The experimental comparison is presented in Fig. \ref{fig:SEED_U}. It shows that the introduced unlabeled source domain ($\mathbb{U}$) is beneficial to the emotion recognition task under different label scarcity conditions and helps the model training process to avoid the overfitting problem, especially when the labeled source domain ($\mathbb{S}$) is extremely small.

\subsection{The effect of hyperparameter settings}
The effect of the hyperparameter settings used in the proposed EEGMatch is also examined.  (1) We change the temperature parameter in the $Sharpen$ function (Eq. \ref{Eq:sharpen}) from 0.1 to 1 with a step of 0.1. The corresponding results are reported in Fig. \ref{fig:hyperparameter} (a). It shows that a change in the temperature parameter would slightly affect the model performance, where the optimal temperature parameter value is 0.2 for the SEED database. (2) We adjust the augmentation size (the ratio of the number of augmented samples to the number of original samples) from 0 to 5 in the experiment, where 0 indicates no data augmentation and 5 refers to fourfold data augmentation. As reported in Fig. \ref{fig:hyperparameter} (b), an increase in the augmentation size could improve the model performance; when the augmentation size reaches 2, the model performs at its optimal. When the augmentation size exceeds 4, it is evident that the EEG-Mixup augmentation begins to negatively impact the model's performance. It indicates that a substantial number of the augmented samples might have an adverse effect on the validity and reliability of the training data. This phenomenon has also been reported by other studies\cite{luo2018aug,luo2020aug}. In Supplementary Materials D, E, and F, we provide more sensitivity analysis on the other hyper-parameters using the same experiment setting ($N=3$), such as the number of trainable parameters, the selection of EEG frequency bands, and the initialization of dynamic thresholds.

\begin{figure}[h]
\begin{center}
\includegraphics[width=0.5\textwidth]{Temp_analysis_aug_analysis_seedYWS.pdf}
\caption{The emotion recognition accuracies (\%) of the proposed EEGMatch at different hyperparameter settings ($N = 3$). (a) The effect of the temperature parameter in the $Sharpen$ function (Eq. \ref{Eq:sharpen}). (b) The effect of the augmentation size in the EEG-Mixup based data augmentation.}
\label{fig:hyperparameter}
\end{center}
\end{figure}
    
\subsection{Topographic analysis of important EEG patterns}
\begin{figure}[h]
\begin{center}
\includegraphics[width=0.5\textwidth]{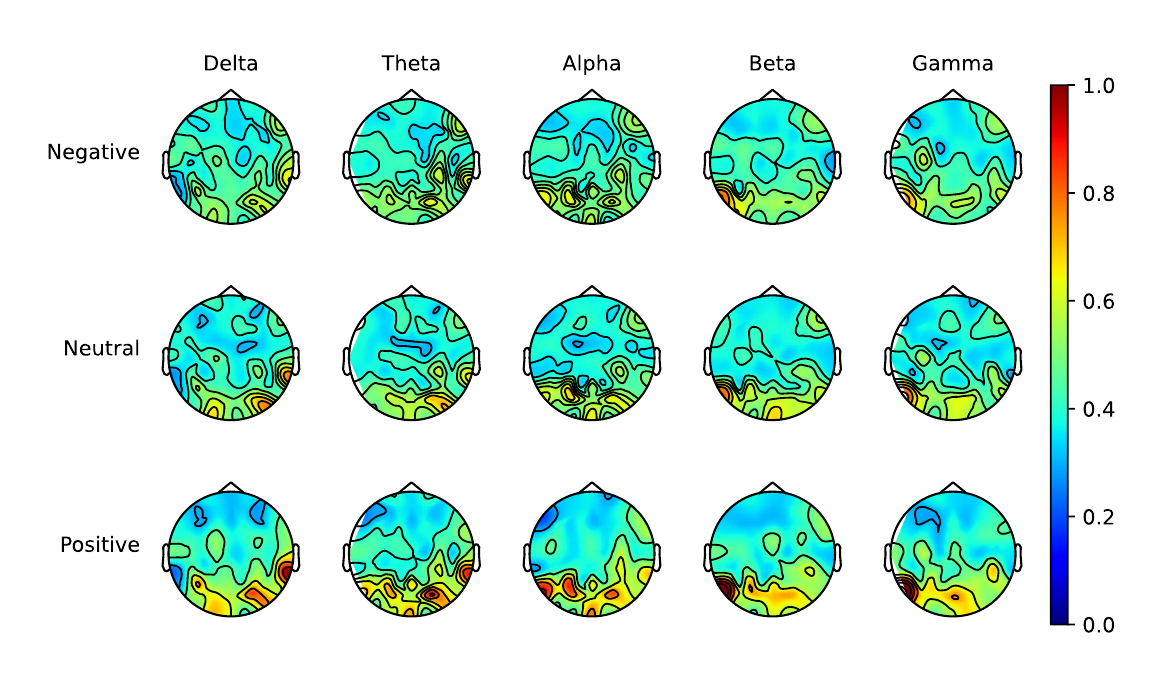}
\end{center}
\caption{Topographic analysis of the mutual information between the EEG patterns and the model predictions.}
\label{fig:topo}
\end{figure}

In the topographic analysis, we identify the important brain patterns for emotion recognition by computing the mutual information between the brain patterns and prediction labels. Specifically, at the $i$th validation round, we have the data input of the target domain, given as $X^t_i$, which is a $M\times310$ matrix. Here, $M$ is the sample size in the target domain, and $310$ refers to the extracted DE features from 62 channels at 5 frequency bands. The corresponding model prediction results are $\hat{Y}^t_i$, with a size of $M\times3$. The columns in $\hat{Y}^t_i$ indicate the prediction probabilities of different emotions (negative, neutral, and positive). Then, we estimate the mutual information between the input features $X^t_i$ and the prediction results $\hat{Y}^t_i$ using the non-parametric method as stated in\cite{kozachenko1987sample,kraskov2004estimating,ross2014mutual}. The obtained mutual information matrix is termed as $I(X^t_i,\hat{Y}^t_i) \in R^{3\times310}$, indicating a quantification of the inherent dependence between the EEG patterns and the model prediction results. $I(X^t_i,\hat{Y}^t_i)$ is further normalized to $[0,1]$, and a larger value refers to a greater informativeness of the EEG patterns to the model prediction at the $i$th validation round. Fig. \ref{fig:topo} shows an average of all the obtained $I(X^t_i,\hat{Y}^t_i)$ across different validation rounds. It is found that the EEG patterns with greater informativeness for emotion recognition are mainly located in the Beta and Gamma frequency bands\cite{zheng2015investigating,song2018eeg} at the parieto-occipital regions\cite{zhong2020eeg,Du2020}.

\subsection{Conclusion}
In this paper, we present a novel semi-supervised learning framework (EEGMatch) for EEG-based cross-subject emotion recognition with incomplete labels. The proposed EEGMatch introduces an efficient EEG-Mixup based data augmentation to increase the sample size for modeling, develops a semi-supervised two-step pairwise learning framework (prototype-wise and instance-wise) for efficient global and local feature learning, and proposes a semi-supervised multi-domain adaptation method to jointly tackle the distribution mismatch problem among multiple domains. The extensive experiments are conducted on two benchmark databases (SEED and SEED-IV) under a practical experimental protocol under various label scarcity conditions. In comparison to the existing popular machine learning and deep learning methods, the experimental results demonstrate the superiority of the proposed EEGMatch in tackling semi-supervised EEG applications with incomplete labels. 

The main performance improvement achieved by the proposed EEGMatch model can be attributed to four factors. (1) The proposed EEG-Mixup based data augmentation method can generate suitable augmented samples for model training, which can help the model avoid the overfitting problem under label scarcity conditions. (2) Through the pseudo label generation with the sharpen function, the proposed semi-supervised prototype-wise pairwise learning could take advantage of both labeled and unlabeled source data and accurately estimate the prototypical representation of each emotion class. (3) The introduced semi-supervised instance-wise pairwise learning can estimate reliable instance-to-instance relationships and project them into a robust feature embedding space for emotion recognition, where the samples from the same class are pulled together and the samples from different classes are pushed apart. (4) The proposed semi-supervised multi-domain adaptation method can simultaneously align the distribution of different domains and help the model predict accurate labels for the samples in $\mathbb{U}$ and $\mathbb{T}$ domains.

Overall, our work presents an efficient and effective semi-supervised learning framework for EEG-based cross-subject emotion recognition. By taking into account the distribution shift, the semi-supervised EEG-based cross-subject emotion recognition performance could be greatly enhanced. We believe that the proposed EEGMatch would open up new possibilities for resolving the label missing problem in EEG tasks, in which high-quality labels are challenging to acquire. In the future, further consideration of the class imbalance issue in modeling is necessary, which commonly happens in real-world applications. Moreover, the stabilization of multi-domain adversarial training poses an ongoing challenge that warrants increased attention in future research efforts.

\newpage
\section*{Appendix A\\Proof of the computation $P$} 
For the total $c$ emotion classes, the supervised prototypical representation $P$ is computed as
\begin{equation}
    P=\left(Y^TY\right)^{-1}Y^Tf\left(X\right),
\end{equation}
where $X=[x_1;x_2,...;x_n]$ and $Y=[y_1;y_2,...;y_n]$ are the labeled source data and the corresponding emotional labels (one-hot representation) in the current mini-batch, respectively. $f\left(\cdot\right)$ is a feature extractor with the parameter $\theta_f$. $f\left(X\right)$ is a $N_B\times M$ feature matrix, where $N_B$ is the batch size for stochastic gradient descent and $M$ is the feature dimensionality. $P=\left[\mu_{1};\mu_{2};...;\mu_{c}\right]$, where $\mu_{c}$ is the prototypical representation for the $c$th emotional category. 

\emph{Proof.}
\begin{equation}
\begin{split}  P=&\left[\mu_{1};\mu_{2};...;\mu_{c}\right]\\=&\left[\frac{1}{N_1}\sum_{i_1}^{N_1}f(x_{i_1});\frac{1}{N_2}\sum_{i_1}^{N_2}f(x_{i_2});...;\frac{1}{N_c}\sum_{i_c}^{N_c}f(x_{i_c})\right]\\=&
    \begin{bmatrix} 
	\frac{1}{N_1} &0&\cdots&0 \\
	0 &\frac{1}{N_2}&\cdots&0\\
	\vdots&\vdots& \ddots&\vdots \\
	0&0&\cdots&\frac{1}{N_c}
	\end{bmatrix}
	\left[\sum_{i_1}^{N_1}f(x_{i_1});...;\sum_{i_c}^{N_c}f(x_{i_c})\right]\\=&
	\begin{bmatrix} 
	{N_1} &0&\cdots&0 \\
	0 &{N_2}&\cdots&0\\
	\vdots&\vdots& \ddots&\vdots \\
	0&0&\cdots&{N_c}
	\end{bmatrix}^{-1}
	Y^{T}f(X)\\=&
	\left(Y^TY\right)^{-1}Y^Tf\left(X\right),
\end{split}
\nonumber
\end{equation}
where $N_j$ ($j \in [1,...,c]$) is the sample size in the $j$th emotion class. $x_{i_j}$ are the EEG samples belonging to the $j$th emotion class.

\section*{Appendix B\\Proof of Theorem 1}
\textbf{Theorem 1:} \emph{Based on the labeled source domain ($\mathbb{S}$), the unlabeled source domain ($\mathbb{U}$), and the target domain ($\mathbb{T}$), the target error $\epsilon_{T}(h)$ with a hypothesis $h$ ($h \in \mathcal{H}$) is bounded as}
\begin{equation}
\begin{split}
\epsilon_{T}(h) &\leq \epsilon_{S}(h)+\min \left\{\mathbb{E}_{\overline{\mathcal{D}}_{T}}\left[\left|F_{\overline{T}}-F_{T}\right|\right], \mathbb{E}_{\mathcal{D}_{T}}\left[\left|F_{T}-F_{\overline{T}}\right|\right]\right\}\\&+\pi_{U}(d_{\mathcal{H}}\left(\mathcal{D}_{S}, \mathcal{D}_{U}\right)+d_{\mathcal{H}}\left(\mathcal{D}_{U}, \mathcal{D}_{T}\right))+\pi_{S}d_{\mathcal{H}}\left(\mathcal{D}_{S}, \mathcal{D}_{T}\right)\\&+\pi_{U}\min \left\{\mathbb{E}_{\mathcal{D}_{S}}\left[\left|F_{S}-F_{U}\right|\right], \mathbb{E}_{\mathcal{D}_{U}}\left[\left|F_{U}-F_{S}\right|\right]\right\},
\end{split}
\nonumber
\end{equation}
\emph{where $\epsilon_{S}(h)$ is the error of the labeled source domain, and $d_{\mathcal{H}}$ represents the $\mathcal{H}$-divergence between the given domains. $F_{S}(x)$, $F_{U}(x)$, and $F_{T}(x)$ are the labeling functions of the labeled source domain, the unlabeled source domain, and the target domain, respectively. $F_{\overline{T}}(x)= \pi_{S} F_{S}(x)+\pi_{U} F_{U}(x)$ is a labeling function for any $x \in Supp(\overline{D}_T)$, which is a weighted summation of $F_{S}(x)$ and $F_{U}(x)$ with the weights of $\pi_S$ and $\pi_U$.}

\emph{Proof.} In the standard domain adaptation, the target error $\epsilon_{T(h)}$ for a given classification hypothesis $h \in \mathcal{H}$ is bounded by\cite{ben2010theory}
\begin{equation}
\begin{split}
\epsilon_{T}(h) &\leq \epsilon_{S}(h)+d_{\mathcal{H} }\left(\mathcal{D}_{S}, \mathcal{D}_{T}\right)+\delta,
\end{split}
\end{equation}
where $\mathcal{D}_{S}$ and $\mathcal{D}_{T}$ refer to the labeled source domain and the target domain, respectively. $\delta=\min \left\{\mathbb{E}_{\mathcal{D}_{S}}\left[\left|F_{S}-F_{T}\right|\right], \mathbb{E}_{\mathcal{D}_{T}}\left[\left|F_{T}-F_{S}\right|\right]\right\}$. However, under the setting of this study, the source domain is composed of both labeled and unlabeled data. If simply applying the standard domain adaptation in this study by only using the labeled source data, the estimation of the target error would not be accurate enough. 

Motivated by the domain projection method introduced in the domain generalization studies\cite{albuquerque2019generalizing,zhao2018adversarial}, we can project the target domain $\mathcal{D}_{T}$ onto the convex hull of the source and compute its "projection" as Eq. \ref{eq:projection}. Based on the given labeled source domain ($\mathcal{D}_{S}$), the unlabeled source domain ($\mathcal{D}_{U}$), and the target domain ($\mathcal{D}_{T}$), the computation is given as

\begin{equation}
\label{eq:projection}    \mathcal{D}_{\overline{T}}=\operatorname{argmin}_{\pi_{S},\pi_{U}} d_{\mathcal{H}}\left[\mathcal{D}_{T},  \pi_{S}\mathcal{D}_{S}+\pi_{U} \mathcal{D}_{U}\right],
\end{equation}
where $\pi_{S}$ and $\pi_{U}$ are belong to $\Delta_{1}$ and  $\pi_{S}+\pi_{U}$ is equal to 1. The labeling function of $\overline{\mathcal{D}}_{T}$ could be represented by a weighted summation of the labeling functions of \textcolor{black}{$\mathcal{D}_{S}$ and $ \mathcal{D}_{U}$}, as
\begin{equation}
    f_{\overline{T}}(x)= \pi_{S} F_{S}(x)+\pi_{U} F_{U}(x),
\end{equation}
\textcolor{black}{where $F_{S}(x)$ and $F_{U}(x)$ are the corresponding labeling functions of $\mathcal{D}_{S}$ and $ \mathcal{D}_{U}$.} Then, the target error is bounded as
\begin{equation}
\label{Eq:appendix_bound}
 \epsilon_{T}(h) \leq 
 \delta_{\overline{T},T}+\epsilon_{\overline{T}}(h)+d_{\mathcal{H} }\left(\mathcal{D}_{\overline{T}}, \mathcal{D}_{T}\right) ,
\end{equation}
where $\delta_{\overline{T},T}=\min \left\{\mathbb{E}_{\mathcal{D}_{\overline{T}}}\left[\left|F_{\overline{T}}-F_{T}\right|\right], \mathbb{E}_{\mathcal{D}_{T}}\left[\left|F_{T}-F_{\overline{T}}\right|\right]\right\}$. The overall source error $\epsilon_{\overline{T}}(h)$ is defined as
\begin{equation}
    \epsilon_{\overline{T}}(h)=\pi_{S}\epsilon_{S}(h)+\pi_{U}\epsilon_{U}(h),
\end{equation}
which is a weighted summation of the labeled source error of $\epsilon_{S}(h)$ and the unlabeled source error of $\epsilon_{U}(h)$. Considering that the label information of the unlabeled source domain $\mathcal{D}_{U}$ is unknown in the training process, we can bound the error $\epsilon_{U}(h)$ as
\begin{equation}
 \epsilon_{U}(h) \leq 
 \delta_{S,U}+\epsilon_{S}(h)+d_{\mathcal{H} }\left(\mathcal{D}_{S}, \mathcal{D}_{U}\right), 
\end{equation}
where $\delta_{S,U}=\min \left\{\mathbb{E}_{\mathcal{D}_{S}}\left[\left|F_{S}-F_{U}\right|\right],\mathbb{E}_{\mathcal{D}_{U}}\left[\left|F_{U}-F_{S}\right|\right]\right\}$. The $\mathcal{H}$-divergence between $\mathcal{D}_{\overline{T}}$ and $\mathcal{D}_{T}$ is given as
\begin{equation}
\begin{split}
d_{\mathcal{H} }\left(\mathcal{D}_{\overline{T}}, \mathcal{D}_{T}\right)=2 \sup _{h \in \mathcal{H}}|\operatorname{Pr}_{x \sim \mathcal{D}_{\overline{T}}}[h(x)=1]\\-\operatorname{Pr}_{x \sim \mathcal{D}_{T}}[h(x)=1]|.
\end{split}
\end{equation}
According to the triangle inequality and the sub-additivity of the $\sup$, the upper-bound of the $\mathcal{H}$-divergence can be written as\cite{albuquerque2019generalizing,zhao2018adversarial}
\begin{equation}
d_{\mathcal{H} }\left(\overline{\mathcal{D}}_{T}, \mathcal{D}_{T}\right)\leq 
\pi_{S}d_{\mathcal{H}}\left(\mathcal{D}_{S}, \mathcal{D}_{T}\right)
+\pi_{U}d_{\mathcal{H}}\left(\mathcal{D}_{U}, \mathcal{D}_{T}\right).
\end{equation}
Through replacing the $\epsilon_{\overline{T}}(h)$ with $\pi_{S}\epsilon_{S}(h)+\pi_{U}\epsilon_{U}(h)$ and replacing the upper-bound $d_{\mathcal{H} }\left(\overline{\mathcal{D}}_{T}, \mathcal{D}_{T}\right)$ with $\pi_{S}d_{\mathcal{H}}\left(\mathcal{D}_{S}, \mathcal{D}_{T}\right)
+\pi_{U}d_{\mathcal{H}}\left(\mathcal{D}_{U}, \mathcal{D}_{T}\right)$ in Eq. \ref{Eq:appendix_bound}, the proof completes.

\section*{Appendix C\\Details about the EEG-Mixup augmentation}
\begin{figure*}[h]
\begin{center}
\subfloat[ EEG-Mixup on the trials with labels (domain $\mathbb{S}$)]{\includegraphics[width=0.5\textwidth]{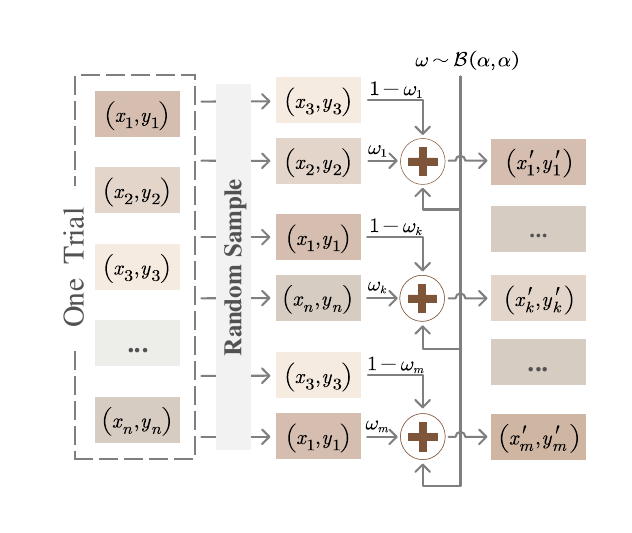}}
\subfloat[ EEG-Mixup on the trials without labels (domains $\mathbb{U}$ and $\mathbb{T}$)]
{\includegraphics[width=0.5\textwidth]{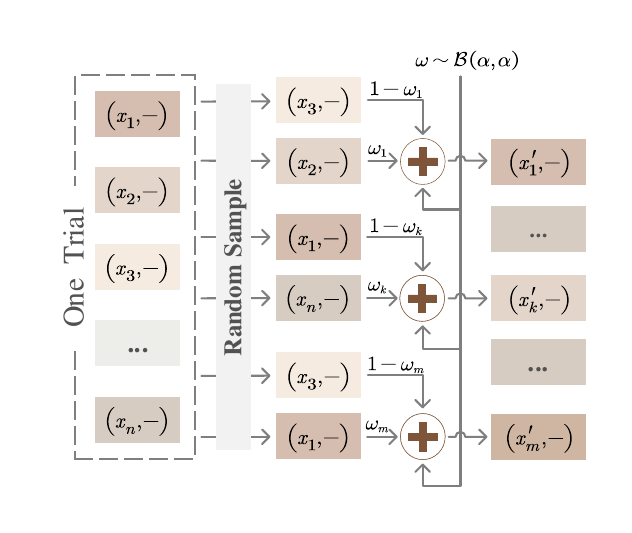}}
\caption{\textcolor{black}{Diagram of the proposed EEG-Mixup augmentation}}
\label{fig:EEGMixup}
\end{center}
\end{figure*}

\textcolor{black}{In this section, we provide a detailed description of the EEG-Mixup augmentation. The main idea of the EEG-Mixup is to fit the
original EEG data in accordance with the independent and identically distributed (IID) assumption and then apply mixup augmentation\cite{zhang2018mixup} to generate augmented data. Specifically, we first sample EEG signal $x_n$ and label $y_n$ from the same trial of the same subject, which follows the joint distribution. As shown in Fig. \ref{fig:EEGMixup} (a), we randomly select two samples from one trial and then generate a new sample ($x_{k}^{'}$ and $y_{k}^{'}$) using the Mixup augmentation. Note that for the trials without labels (domains $\mathbb{U}$ and $\mathbb{T}$), we only use the EEG signal for augmentation, which is presented in Fig. \ref{fig:EEGMixup} (b). To illustrate the EEG-Mixup from an intuitive perspective, we also provide the corresponding pseudo-code in Algorithm S1.}

\begin{algorithm*}[]
  \caption{The training algorithm of EEGMatch}
  \label{alg::conjugateGradient}
  \begin{algorithmic}[1]
    \Require
\renewcommand{\algorithmicrequire}{\textbf{}}
    \Require - emotion class $c$, max iteration $Maxepoch$, batch size $N_B$, initial threshold values $\tau_h^0$ and $\tau_l^0$;
    \Require - labeled source data $\{X_s,Y_s\}$, unlabeled source data $\{X_u\}$, unknown target data $\{X_t\}$;
    \Require - random initialization of prototype representation $P_{s^{*}}$, bilinear transformation matrix $B$, feature extractor $f(\cdot)$, and discriminator $d(\cdot)$;
    \Ensure $P_{s^{*}}$, $B$, $f(\cdot)$, $d(\cdot)$; 
    \Require  \textcolor{gray}{\#\#All the equations used below can be found in Section III of the main manuscript.\#\#}
    \Require \textcolor{gray}{\#\# EEG-Mixup based data augmentation \#\#}
          \State generate the augmented data $\{\mathcal{A}(X_s,Y_s)\}$, $\{\mathcal{A}(X_u)\}$, $\{\mathcal{A}(X_t)\}$ using Eq. (1) and (2);
          \State update $\{X_s,Y_s\}$ by concatenating $\{X_s,Y_s\}$ and the augmented data $\{\mathcal{A}(X_s,Y_s)\}$;
          \State update $\{X_u\}$ by concatenating $\{X_u\}$ and the augmented data $\{\mathcal{A}(X_u)\}$;
          \State update $\{X_t\}$ by concatenating $\{X_t\}$ and the augmented data $\{\mathcal{A}(X_t)\}$;
    \For {$1$ to $Maxepoch$} 
         \For {$b$ = 1 to $N_B$}
          \State sample labeled source domain data $\{X_s^b,X_s^b\}$ from the updated $\{X_s,Y_s\}$; 
          \State sample unlabeled source domain data $\{X_u^b\}$ from the updated $\{X_u\}$;
          \State sample unlabeled target domain data $\{X_t^b\}$ from the updated $\{X_t\}$;
         \Require \textcolor{gray}{\#\# semi-supervised two-step pairwise learning \#\#}
         \State  $Cls\left(f\left(X_{u}^{b}\right)\right)=softmax\left(f\left(X_{u}^{b}\right)^{T} B P_{s^{*}}\right)$; 
         \State $\widehat{Y}_{u}^{b}=Sharpen\left(Cls\left(f\left(X_{u}^{b}\right)\right)\right)$; \textcolor{gray}{\qquad\qquad\qquad\qquad\qquad\qquad\qquad\qquad\qquad\qquad\qquad\qquad// pseudo labels of $\mathbb{U}$ }
         \State update $X^{s^*}=\left[X_s^{b};X_u^{b}\right]$, $Y^{s^*}=\left[Y_s^{b};\hat{Y}_{u}^{b}\right]$; 
         \State  calculate $P_{s^{*}}=\left((Y^{s^*})^T(Y^{s^*})\right)^{-1}(Y^{s^*})^Tf\left(X^{s^*}\right)$; \textcolor{gray}{\qquad\qquad\qquad\qquad\qquad// prototypical representation}
         \State compute $R^s_p$ with $X^{s^*}$ using Eq. (12) and compute $R^t_p$ with $X_t^b$ using Eq. (15); \textcolor{gray}{// instance-to-instance similarity matrix}
         \State compute $R^s$ with $Y^{s^*}$ using Eq. (13) and compute $R^t$ using Eq. (16); \textcolor{gray}{\qquad// groundtruth similarity matrix}
         \State compute $\mathcal{L}_{pair}^s$ using Eq. (14), and compute $\mathcal{L}_{pair}^t$ using Eq. (17) and (18);
         \State compute $\mathcal{L}_{pair}$ using Eq. (20);  \textcolor{gray}{// instance-wise pairwise learning loss computation based on $\mathcal{L}_{pair}^s$ and $\mathcal{L}_{pair}^t$}
         \State update $\tau_h$ and $\tau_l$ using Eq. (19); \textcolor{gray}{\qquad\qquad\qquad\qquad\qquad\qquad\qquad\qquad\qquad\qquad\qquad// dynamic thresholds}
         \Require \textcolor{gray}{\#\# semi-supervised multi-domain adaptation \#\#}
         \State compute $\mathcal{L}_{disc}$ using Eq. (25); \textcolor{gray}{\qquad\qquad\qquad// semi-supervised multi-domain adaptation loss computation}
         \State compute $\mathcal{L}$ using Eq. (26); \textcolor{gray}{\qquad\qquad\qquad\qquad// the final objective loss computation based on $\mathcal{L}_{pair}$ and $\mathcal{L}_{disc}$}
         \State gradient back-propagation;
         \State update network parameters;
    \EndFor
\EndFor
\end{algorithmic}
\end{algorithm*}

\begin{algorithm*}[]
  \caption{The pseudo-code of EEG-Mixup}
  \label{alg::EEGMixup}
  \color{black}
  \begin{algorithmic}[1]
    \Require
\renewcommand{\algorithmicrequire}{\textbf{}}
    \Require - labeled source data $\{X_s,Y_s\}$, 
    unlabeled source data $\{X_u\}$, target data $\{X_t\}$;
    \Require - The number of source domain subjects $Sub$, the number of labeled trials in the source domain $N$, the number of unlabeled trials in the source domain $M-N$;
    \Require - The number of target domain subjects $Sub_t$, the number of unlabeled trials in the target domain $M$;
    \Require - Augmentation ratio $ratio$;
    \Ensure augmented data $\{\mathcal{A}(X_s,Y_s)\}$, $\{\mathcal{A}(X_u)\}$, $\{\mathcal{A}(X_t)\}$; 
    \Require \textcolor{gray}{\#\# EEG-Mixup on the labeled source domain \#\#}
    \For {$p_s$ = 1 to $Sub$} 
         \For {$q_s$ = 1 to $N$}
             \State collect $D_{p_s,q_s}^{s}=\left\{\left(x_{n}^s,y_{n}^s\right)\right\}_{n=1}^{N^s_{p_s,q_s}}$ from $\{X_s,Y_s\}$;
            \State \textcolor{gray}{//$N^s_{p_s,q_s}$ is the number of samples from $p_s$ subject at $q_s$ trial.}
            \For {1 to $ ratio \times N^s_{p_s,q_s}$ }
             \State randomly select two samples $\bigl(x_{i}^s,y_{i}^s\bigr)$ and $\bigl(x_{j}^s,y_{j}^s\bigr)$;
             \State sample $\omega$ from distribution $\operatorname{Beta}(\alpha, \alpha)$
             \State generate new sample by $x_{z}^s=\omega x_i^s+(1-\omega)x_j^s$, $y_{z}^s=\omega y_i^s+(1-\omega)y_j^s$;
             \State add $\left(x_{z}^s,y_{z}^s\right)$ to $\mathcal{A}(X_s,Y_s)$
             \EndFor
    \EndFor
\EndFor

\Require \textcolor{gray}{\#\# EEG-Mixup on the unlabeled source domain \#\#}
    \For {$p_u$ = 1 to $Sub$} 
         \For {$q_u$ = 1 to $M-N$}
             \State collect $D_{p_u,q_u}^{u}=\left\{\left(x_{n}^u,y_{n}^u\right)\right\}_{n=1}^{N^u_{p_u,q_u}}$ from $\{X_u\}$;
            \State \textcolor{gray}{//$N^u_{p_u,q_u}$ is the number of samples from $p_u$ subject at $q_u$ trial.}
            \For {1 to $ ratio \times N^u_{p_u,q_u}$ }
             \State randomly select two samples $x_{i}^u$ and $x_{j}^u$;
             \State sample $\omega$ from distribution $\operatorname{Beta}(\alpha, \alpha)$
             \State generate new sample by $x_{z}^u=\omega x_i^u+(1-\omega)x_j^u$;
             \State add $x_{z}^u$ to $\mathcal{A}(X_u)$
             \EndFor
    \EndFor
\EndFor

\Require \textcolor{gray}{\#\# EEG-Mixup on the target domain \#\#}
    \For {$p_t$ = 1 to $Sub_t$} 
         \For {$q_t$ = 1 to $M$}
             \State collect $D_{p_t,q_t}^{t}=\left\{\left(x_{n}^t,y_{n}^t\right)\right\}_{n=1}^{N^t_{p_t,q_t}}$ from $\{X_t\}$;
            \State \textcolor{gray}{//$N^t_{p_t,q_t}$ is the number of samples from $p_t$ subject at $q_t$ trial.}
            \For {1 to $ ratio \times N^t_{p_t,q_t}$ }
             \State randomly select two samples $x_{i}^t$ and $x_{j}^t$;
             \State sample $\omega$ from distribution $\operatorname{Beta}(\alpha, \alpha)$
             \State generate new sample by $x_{z}^t=\omega x_i^t+(1-\omega)x_j^t$;
             \State add $x_{z}^t$ to $\mathcal{A}(X_t)$
             \EndFor
    \EndFor
\EndFor
\end{algorithmic}
\end{algorithm*}

\section*{Appendix D\\Effect of the EEG bands used for model training}
\textcolor{black}{In this section, we report the performance of the proposed EEGMatch in the individual
Delta ($\delta$), Theta ($\theta$), Alpha ($\alpha$), Beta ($\beta$), and Gamma ($\gamma$) bands to investigate their impacts on model performance. For simplicity, we conduct our experiments on the SEED database and set the number of labeled trials as three for each source subject ($N=3$). Table \ref{tab:band} reports the model performance on 15 target subjects using features extracted from various combinations of EEG frequency bands. The results show that the Beta and Gamma bands achieve better recognition results than the other three bands in most subjects. It confirms our findings that Beta and Gamma bands contain the major information for emotion recognition. Furthermore, it is evident that the model trained using all EEG bands demonstrates superior performance when compared to other combination strategies. This phenomenon suggests that a fusion of all EEG bands is essential for achieving optimal performance in our model.  }
\begin{table*}[h]
\setlength{\tabcolsep}{0.45em}
\begin{center}
\caption{The mean accuracies (\%) of the proposed EEGMatch on the SEED database using varying EEG bands.}
\label{tab:band}
\scalebox{1}{
\color{black}
\begin{tabular*}{\hsize}{@{}@{\extracolsep{\fill}}lcccccccccccccccc@{}}
\toprule
Band & sub1 & sub2 & sub3 & sub4 & sub5 & sub6 & sub7 & sub8 & sub9 & sub10 & sub11 & sub12 & sub13 & sub14 & sub15 & Average\\
\midrule
$\delta$ band&58.99&55.24&50.53&70.95&77.22&66.29&77.34&65.20&64.29&78.43&72.16&61.91&58.37&71.83&67.01&66.38$\pm$11.28
\\
$\theta$ band&79.29&74.57&44.41&64.26&50.79&47.58&39.54&54.51&59.51&65.35&53.74&49.21&55.72&43.11&67.74&56.62$\pm$08.11
\\
$\alpha$ band&55.12&44.76&50.23&48.79&68.01&70.18&32.59&61.46&72.15&52.51&56.71&52.44&50.77&55.36&66.41&55.83$\pm$10.19
\\
$\beta$ band&72.16&64.11&58.45&68.01&67.12&69.98&68.23&66.38&73.16&68.79&83.53&60.28&82.41&68.01&95.79&71.09$\pm$09.29
\\
$\gamma$ band  &85.33&60.81&54.39&50.24&67.94&75.78&59.87&85.56&68.89&71.33&91.25&73.48&79.49&85.68&78.11&72.54$\pm$11.85
\\
$\beta$,$\gamma$ bands &83.74 &77.78 &56.12	&90.39	&72.86	&76.93	&87.80	&85.94	&86.38	&82.85	&86.26	&85.59	&80.49	&75.31	&84.29 & 80.85$\pm$08.18
\\
$\beta$,$\gamma$,$\delta$ bands &90.07	&87.95	&78.69	&61.22	&85.21	&92.16	&83.82	&83.44	&81.56	&93.13	&79.19	&73.36	&76.75	&100 &100 &84.43$\pm$09.85\\
$\beta$,$\gamma$,$\theta$ bands &83.32	&76.90	&64.70	&74.19	&87.24	&66.15	&88.16	&87.95	&81.08	&85.65	&93.13	&100 &73.78	&94.05&93.93 &83.35$\pm$10.10\\
$\beta$,$\gamma$,$\alpha$ bands &90.75&63.79&60.81&80.49&71.59&78.43&66.00&83.79&70.53&83.12&92.78&93.93&83.29&85.21&93.08&79.84$\pm$10.63\\
All bands&100.00&79.76&80.55&78.26&78.64&88.42&96.94&71.15&90.48&88.80&93.37&83.35&90.78&95.93&83.82&\textbf{86.68$\pm$07.89}\\
\bottomrule
\end{tabular*}
}
\end{center}
\end{table*}
\section*{Appendix E\\Effect of the number of trainable parameters}
\textcolor{black}{The number of trainable parameters is an important factor that determines the model computation complexity and performance. The computation complexity of the EEGMatch can be estimated by computing the number of trainable parameters in its feature extractor $f(\cdot)$, domain discriminator $d(\cdot)$, and bilinear transformation matrix $B$. For the feature extractor $f(\cdot)$, it is designed as 310 neurons-$H$ neurons-Relu activation-$H$ neurons-Relu activation-$H$ neurons. For the domain discriminator $d(\cdot)$, it is designed as $H$ neurons-$H$ neurons-Relu activation-dropout layer-$H$ neurons-3 neurons-Softmax activation. The size of matrix $B$ is $H \times H$. $H$ refers to the number of neurons in the hidden layers. Consequently, the number of trainable parameters $Num$ in the EEGMatch can be computed as, 
\begin{equation}
    Num=(310H+H^2)+(H^2+3H)+H^2=313H+3H^2 ,
\end{equation}
The higher the number of trainable parameters, the higher the computation complexity. To investigate its effect on the model performance, we adjust the value of $H$ and report the performance of the EEGMatch in Table \ref{tab:complexity}. The experiments are conducted on the SEED database with the three labeled trials ($N=3$). The results indicate that the optimal $H$ value is 64, which achieves the best classification performance. Additionally, it demonstrates the second-lowest computation complexity, indicating a favorable balance between performance and complexity. }
\begin{table*}[h]
\setlength{\tabcolsep}{0.45em}
\begin{center}
\caption{The mean accuracies (\%) of the proposed EEGMatch on the SEED database.}
\label{tab:complexity}
\scalebox{1}{
\color{black}
\begin{tabular*}{\hsize}{@{}@{\extracolsep{\fill}}lcccccccccccccccc@{}}
\toprule
Complexity & sub1 & sub2 & sub3 & sub4 & sub5 & sub6 & sub7 & sub8 & sub9 & sub10 & sub11 & sub12 & sub13 & sub14 & sub15 & Average\\
\midrule
$H=32$&93.48&93.75&76.43&69.79&93.69&80.64&81.67&67.77&85.98&91.96&97.26&100&74.89&90.71&100&86.54$\pm$10.34
\\
\textbf{$H=64$}&100.00&79.76&80.55&78.26&78.64&88.42&96.94&71.15&90.48&88.80&93.37&83.35&90.78&95.93&83.82&\textbf{86.68$\pm$07.89}\\
$H=128$&96.17&78.02&87.09&64.82&77.34&94.55&84.86&89.09&94.76&83.00&90.75&91.84&78.79&93.28&90.93&86.35$\pm$08.35
\\
$H=256$&90.13&83.26&80.58&82.79&94.19&86.18&83.79&85.27&85.53&88.51&90.60&74.72&71.36&90.60&89.22&85.12$\pm$05.92\\

\bottomrule
\end{tabular*}
}
\end{center}
\end{table*}
\section*{Appendix F\\Sensitivity analysis of the dynamic thresholds}
\begin{figure*}[h]
\begin{center}
\includegraphics[width=1\linewidth]{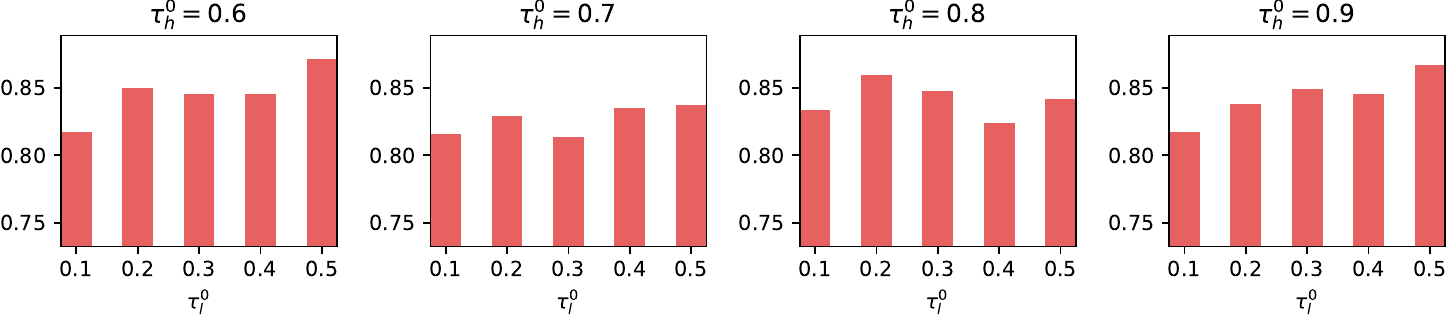}
\caption{\textcolor{black}{The mean accuracies (\%) of the proposed EEGMatch on target domains under different initial thresholds.}}
\label{fig:thres_bar}
\end{center}
\end{figure*}

\textcolor{black}{To ensure correct optimization during model training, we employ two dynamic thresholds ( upper threshold $\tau_h$ and lower threshold
$\tau_l$) to exclude invalid paired samples. In this section, we use a grid-search method to investigate
the impact of initial values of the thresholds on model performance. Specifically, we fix the initial upper threshold $\tau_h^{0}$ and adjust the initial lower threshold $\tau_l^{0}$ from 0.1 to 0.5 with a step of 0.1. For simplicity, we conduct our experiments on the SEED database and set the number of labeled trials as three for each source subject ($N=3$). As illustrated in Fig. \ref{fig:thres_bar}, we can observe that the performance of the proposed EEGMatch is relatively insensitive to the initialization of the dynamic threshold. On the other hand, $\{\tau_h^{0},\tau_l^{0}\}=\{0.9,0.5\}$ and $\{\tau_h^{0},\tau_l^{0}\}=\{0.6,0.5\}$ are two optimal settings that can maximize the model performance.
Considering that a high upper threshold can reduce the confirmation bias\cite{sohn2020fixmatch,arazo2020pseudo}, we initialize the upper threshold as $\tau_h^{0}=0.9$ and the lower threshold as $\tau_l^{0}=0.5$ by default.}
\section*{Appendix G\\Effect of different learning rates}
\begin{figure*}[h]
\begin{center}
\subfloat[learning rate = 0.01]{\includegraphics[width=0.33\linewidth]{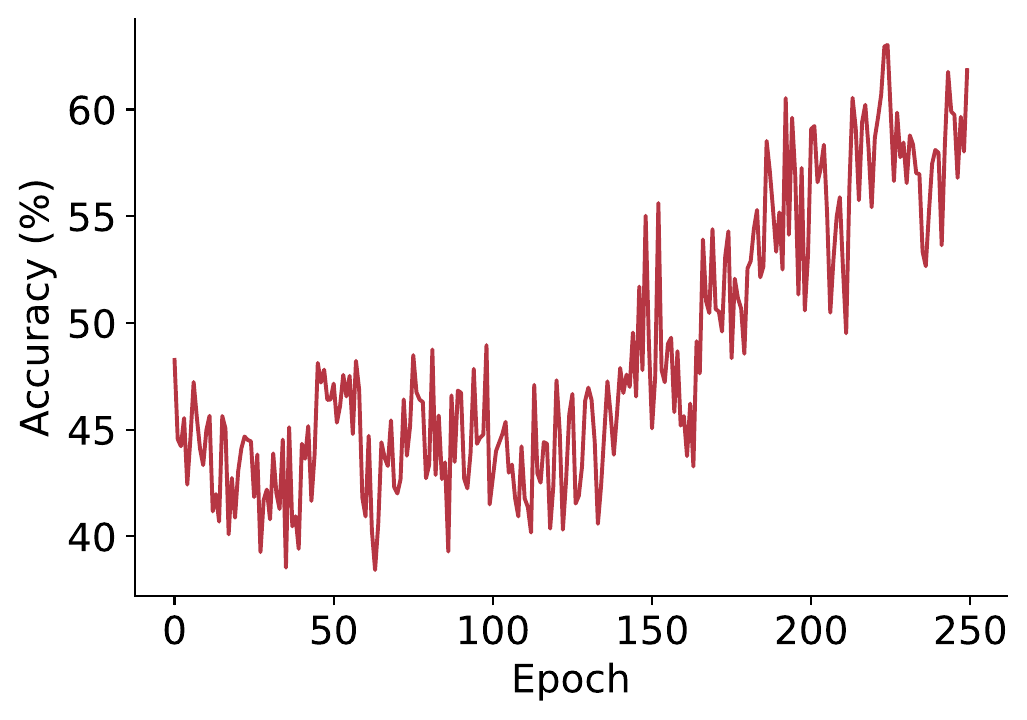}}
\subfloat[learning rate = 0.001]{\includegraphics[width=0.33\linewidth]{target_acc_curve.pdf}}
\subfloat[learning rate = 0.0001]{\includegraphics[width=0.33\linewidth]{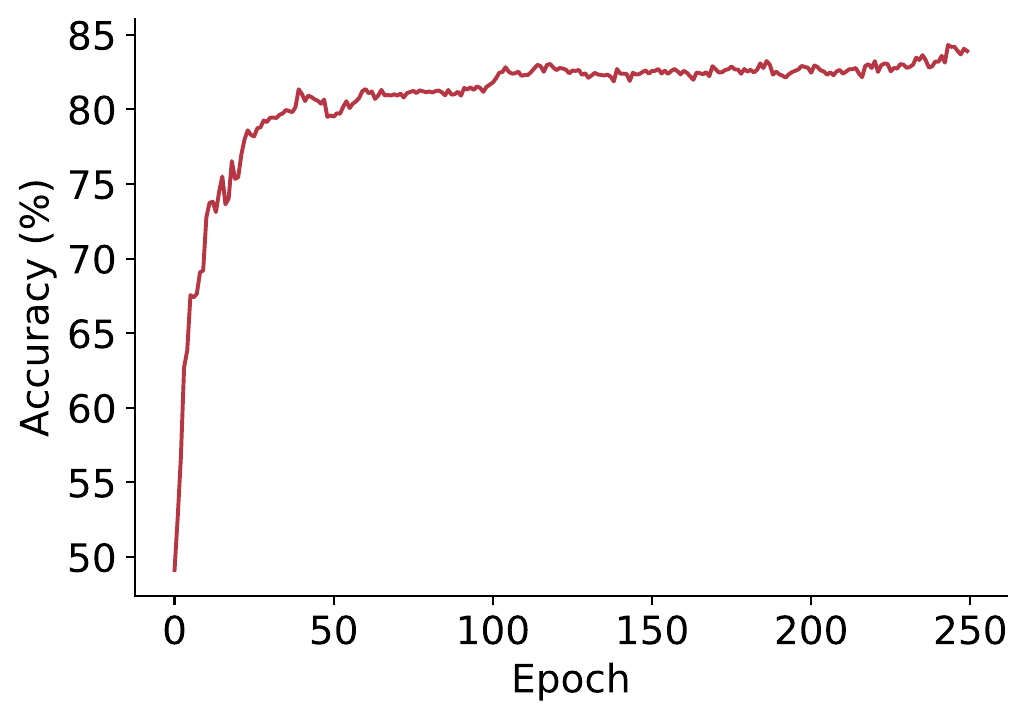}}
\caption{\textcolor{black}{The mean accuracies (\%) of the proposed EEGMatch under different learning rates.}}
\label{fig:curve_lr}
\end{center}
\end{figure*}

\textcolor{black}{In this section, we examine the effect of learning rates on the performance of the proposed EEGMatch. Specifically, we adjust the learning rates in the range of 0.01 to 0.0001 and evaluate the model's performance on the SEED database. Note that the number of labeled trials is set as twelve for each source subject ($N=12$). The accuracy of the EEGMatch on the target domain is recorded during the training process and illustrated in Fig. \ref{fig:curve_lr}. The result demonstrates that small learning rates (0.001 and 0.0001) can stabilize the domain adversarial training and improve the model performance on the target domain. However, we also observe that an excessively small learning rate of 0.0001 would hinder the model's ability to find the optimal parameters, resulting in decreased performance. Consequently, we set the default learning rate of the EEGMatch as 0.001 to jointly consider both optimal model performance and training stability.   }

\section*{Appendix H\\Effect of the number of source subjects}
\begin{figure}[h]
\begin{center}
\includegraphics[width=1\linewidth]{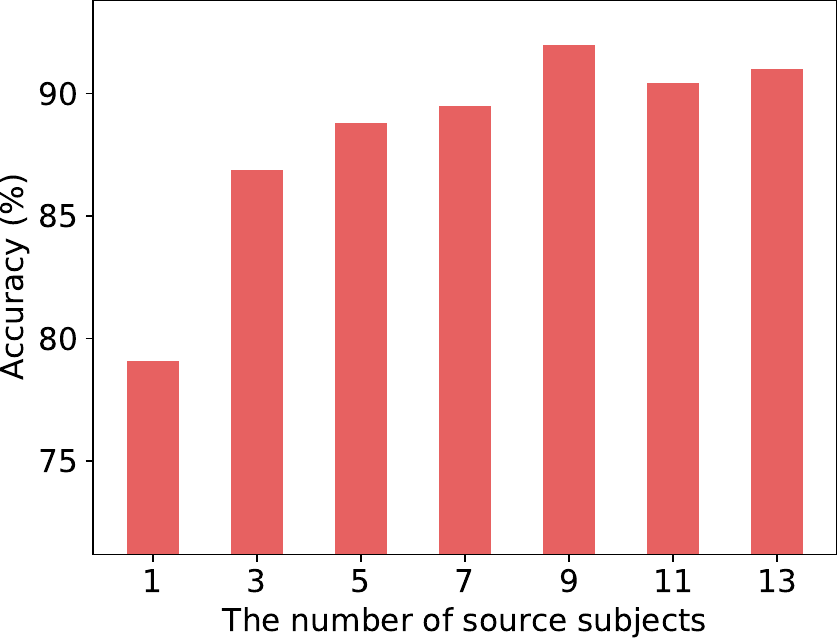}
\caption{\textcolor{black}{The mean accuracies (\%) of the proposed EEGMatch under different number of source subjects.}}
\label{fig:source_num}
\end{center}
\end{figure}

\textcolor{black}{The number of source subjects used for transfer learning is an important factor for model performance\cite{LiJDA2020,peng2022joint}.  In this section, we explore the impact of the number of source subjects, denoted as $M_s$, on the performance of our proposed EEGMatch. Specifically, we adjust the number of source subjects $M_s$ from 1 to 13 with a step of 2 and evaluate the performance of the proposed EEGMatch on the SEED database. In order to ensure that the number of source domain samples is not much smaller than the target domain samples, we set the number of
labeled trials as twelve for each source subject. The mean performance across 15 target subjects of the SEED database is presented in Fig. \ref{fig:source_num}. The results show that the model performance remains stable when $M_s\geq3$, which indicates its robustness across a broader range of source subject quantities. On the other hand, we can determine the minimum number of subjects needed for the model to achieve a specified level of performance. For example, achieving an accuracy exceeding 85\% requires a minimum of 3 subjects, while surpassing an accuracy rate of 90\% necessitates a minimum of 9 subjects.” }
\section*{Appendix I\\Feature visualization on the SEED-IV and SEED-V databases}
\begin{figure*}[h]
\begin{center}
\subfloat[]{\includegraphics[width=0.33\linewidth]{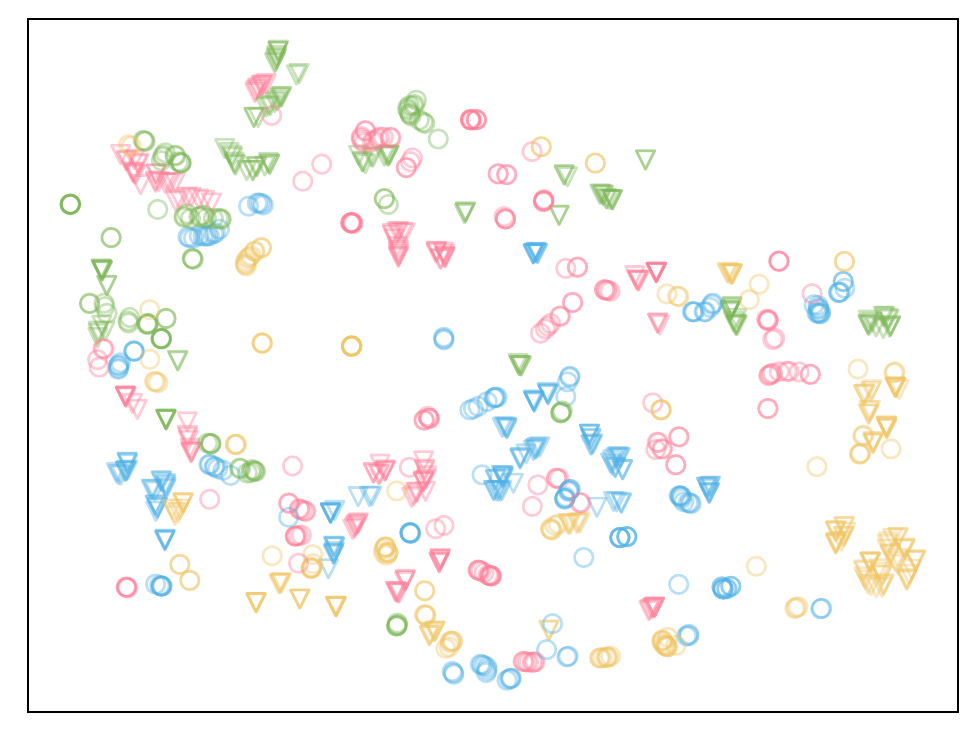}}
\subfloat[]{\includegraphics[width=0.33\linewidth]{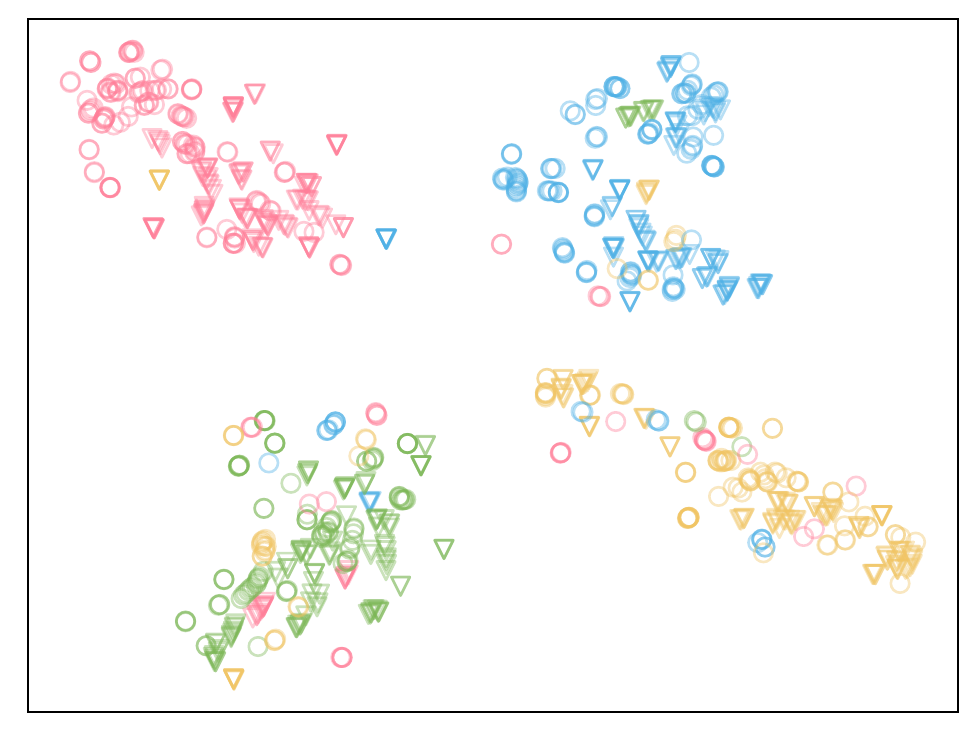}}
\subfloat[]{\includegraphics[width=0.33\linewidth]{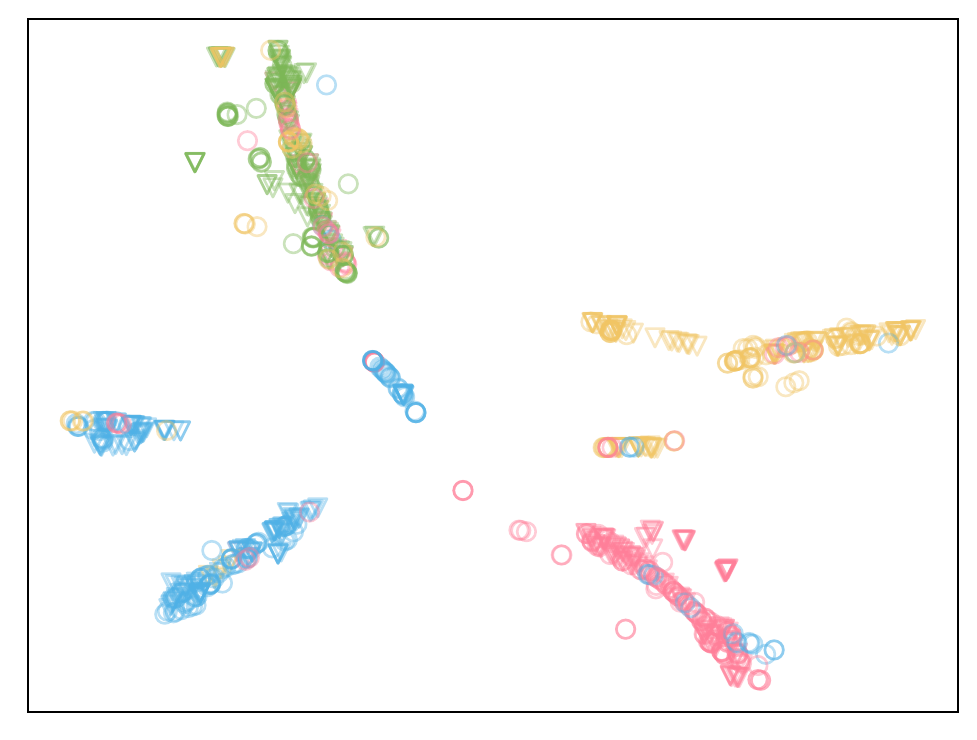}}
\caption{\textcolor{black}{A visualization of the learned feature representations (a) before training, (b) at the training epoch of 50, and (c) in the final model. Here, the circle, asterisk and triangle represent the labeled source domain ($\mathbb{S}$), the unlabeled source domain ($\mathbb{U}$), and the target domain ($\mathbb{T}$). The pink, blue, orange, and green colors indicate happy, sad, fear, and neural emotions. The number of labeled trials is set as $N=20$.}}
\label{fig:tsne_iv}
\end{center}
\end{figure*}

\begin{figure*}[h]
\begin{center}
\subfloat[]{\includegraphics[width=0.33\linewidth]{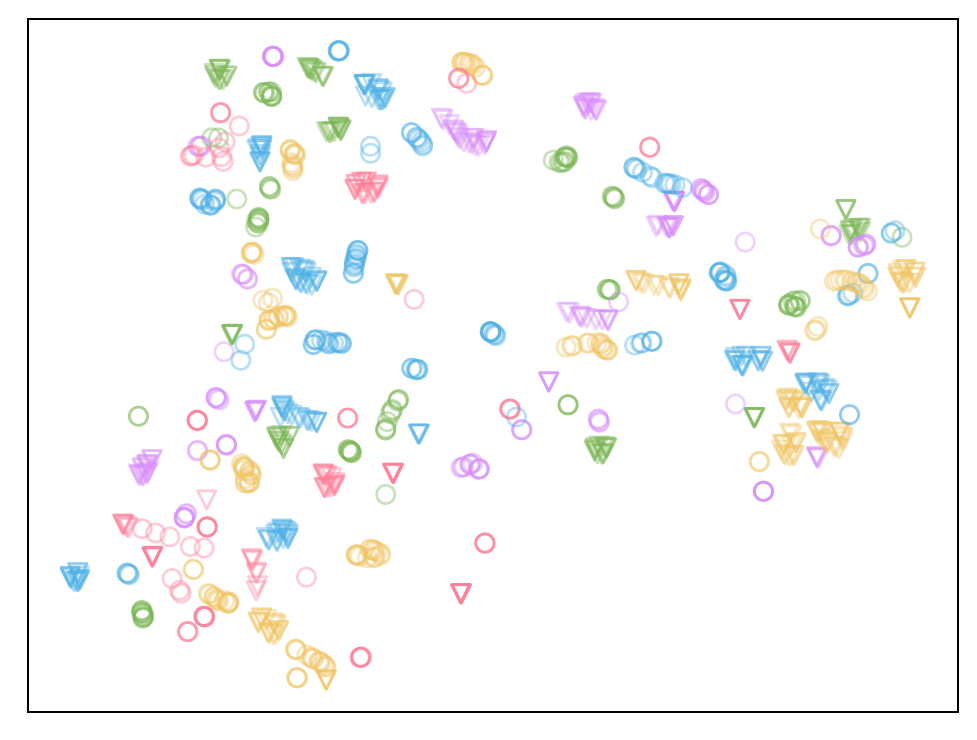}}
\subfloat[]{\includegraphics[width=0.33\linewidth]{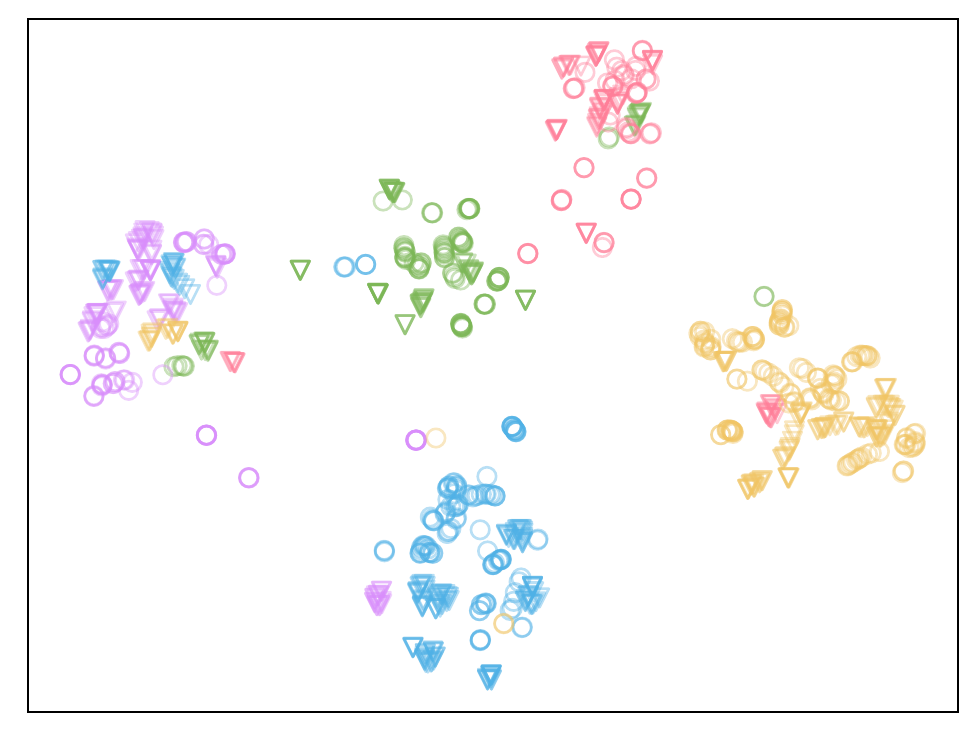}}
\subfloat[]{\includegraphics[width=0.33\linewidth]{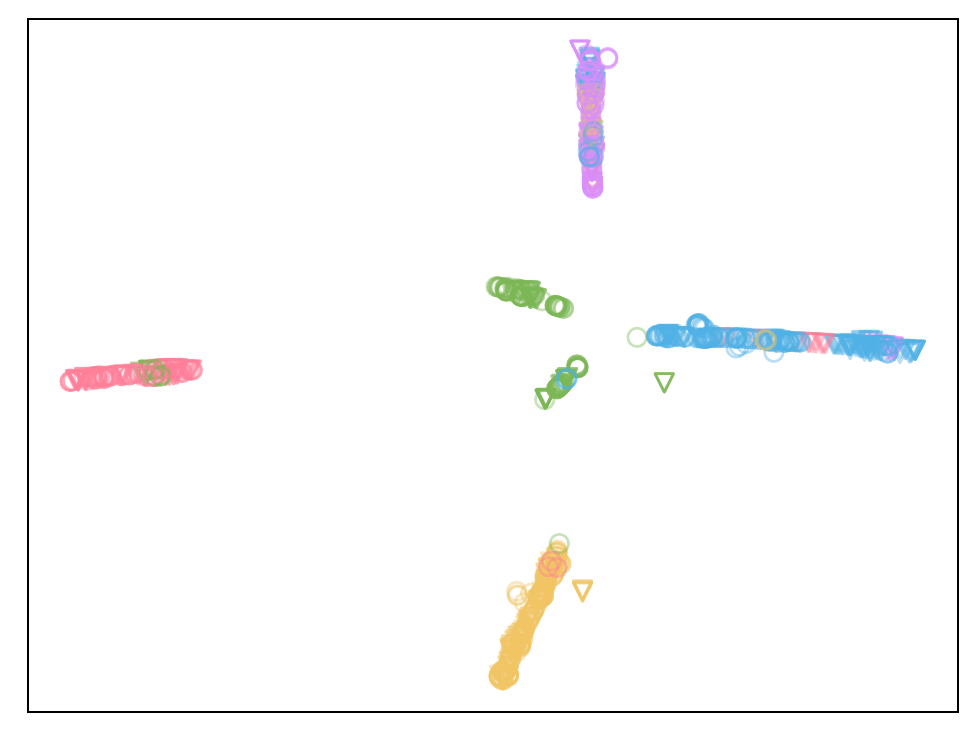}}
\caption{\textcolor{black}{A visualization of the learned feature representations (a) before training, (b) at the training epoch of 50, and (c) in the final model. Here, the circle, asterisk and triangle represent the labeled source domain ($\mathbb{S}$), the unlabeled source domain ($\mathbb{U}$), and the target domain ($\mathbb{T}$). The pink, blue, orange, purple, and green colors indicate happy, sad, neural, fear, and disgust emotions. The number of labeled trials is set as $N=10$.}}
\label{fig:tsne_v}
\end{center}
\end{figure*}

\textcolor{black}{We employ the t-SNE algorithm\cite{2008Tsne} to visualize the learned feature representations from the SEED-IV and SEED-V databases. The results are illustrated in Fig. \ref{fig:tsne_iv} and Fig. \ref{fig:tsne_v}. To enhance clarity, we assign different colors to the samples with different emotions and assign different shapes to distinguish samples from different domains. The results on the two databases demonstrate that EEGMatch reduces the distribution discrepancy among the three domains throughout the training process. Additionally, it progressively separates the samples with different emotions to minimize the emotion classification error. In summary, the results provide supplementary evidence to support the efficacy of the proposed EEGMatch in EEG-based emotion recognition under a semi-supervised learning framework. }

\section*{Appendix J\\Results on the real-world database.}
\textcolor{black}{To further validate the feasibility of the proposed EEGMatch, we extend the application of EEGMatch on a real-world clinical EEG database. This database was obtained from the Hospital Universiti Sains Malaysia (HUSM), featuring patients diagnosed with depression \cite{mumtaz2016mdd}. It includes 27 healthy subjects (38.28$\pm$15.64 years old) and 29 subjects diagnosed with Major Depressive Disorder (MDD) (40.33$\pm$12.86 years old). Their categorization adheres to the international diagnostic criteria for depression outlined in the Diagnostic and Statistical Manual-IV (DSM-IV). The EEG data was collected using the Brain Master Discovery amplifier, with 19 EEG electrodes and a sampling rate of 256Hz. In the preprocessing, bandpass filtering ranging from 0.5 to 70 Hz and a notch filter at 50 Hz were conducted. The experimental protocol involved recording EEG signals from participants during both closed-eye (EC) and open-eye (EO) conditions, with each session lasting 5 minutes. During the open-eye condition, participants were instructed to relax while minimizing eye movements to ensure the quality of the collected data.}

\textcolor{black}{In this study, we employ a cross-subject validation protocol to evaluate the performance of the proposed EEGMatch in handling real-world clinical data. To ensure a robust assessment, we implement a ten-fold cross-validation strategy to compute mean performance and standard deviations. In each validation round, 9 subsets are used as training data, while the remaining 1 subset serves as testing data. This process is iterated 10 times, ensuring that each subset is used as testing data once. To evaluate the model performance under label scarcity conditions, we divided the training subsets into $\mathbb{S}$ and $\mathbb{U}$ domains based on a specified ratio. For instance, if the  $\mathbb{S}: \mathbb{U}$ ratio is $1:1$, 50\% of subjects in the training sets are labeled ($\mathbb{S}$) while the remaining 50\% are unlabeled ($\mathbb{U}$). We adjust the ratio from $1:2$ to $2:1$ and present the results in Table \ref{tab:MDDdata}. Here, the sample balance of the two groups in three domains is considered. The results show that EEGMatch attains an average accuracy of 65.35$\pm$10.07 in the EC experiment, with a 1.33\% improvement compared to the existing literature. In the EO experiment, EEGMatch achieves an average accuracy of 67.08$\pm$09.24, indicating a 2.39\% enhancement compared to other models. In summary, the results demonstrate the efficacy of EEGMatch in alleviating the label-missing problem in a real-world database.}

\begin{table*}[h]
\setlength{\tabcolsep}{0.45em}
\begin{center}
\color{black}
\caption{\textcolor{black}{Experimental results on the MDD database using cross-subject validation protocol with incomplete labels. EC: closed-eye; EO: open-eye.}}
\label{tab:MDDdata}
\scalebox{1}{
\begin{tabular*}{\hsize}{@{}@{\extracolsep{\fill}}ccccccc@{}}
\toprule
Session & $\mathbb{S} : \mathbb{U}$ & MixMatch*\cite{berthelot2019mixmatch} & AdaMatch*\cite{berthelot2021adamatch} & FlexMatch*\cite{zhang2021flexmatch} & SoftMatch*\cite{chen2023softmatch} & \textbf{EEGMatch}\\ 
\midrule
\multirow{3}{*}{\textbf{EC}}&
$2:1$ & 66.56$\pm$06.81 & 66.60$\pm$08.72 & 66.43$\pm$06.50 & 66.63$\pm$03.98 & \textbf{71.32$\pm$08.16} \\\hspace*{\fill}&
$1:1$ &	\textbf{63.48$\pm$07.06} & 62.57$\pm$09.34 & 62.12$\pm$06.73 & 62.24$\pm$06.18 & 61.85$\pm$10.98 \\\hspace*{\fill}&
$1:2$ &	62.01$\pm$07.52 & 61.11$\pm$09.61 & 61.74$\pm$07.06 & 60.19$\pm$09.35 & \textbf{62.88$\pm$07.94} \\
\midrule
\multicolumn{2}{l}{Average Performance}   & 64.02$\pm$07.38 & 63.43$\pm$09.52 & 63.43$\pm$07.09 &  63.02$\pm$07.37 & \textbf{65.35$\pm$10.07} \\
\midrule
\specialrule{0em}{1.5pt}{1.5pt}
\midrule
\multirow{3}{*}{\textbf{EO}}&
$2:1$ & 65.47$\pm$07.81 & 64.54$\pm$08.27 & 61.48$\pm$07.42 & 61.18$\pm$06.99 & \textbf{67.71$\pm$08.48} \\\hspace*{\fill}&
$1:1$ &	65.65$\pm$04.83 & 64.28$\pm$06.05 & 63.52$\pm$09.36 & 63.64$\pm$09.38 & \textbf{66.25$\pm$09.94} \\\hspace*{\fill}&
$1:2$ &	62.95$\pm$06.93 & 61.70$\pm$08.06 & 63.83$\pm$09.53 & 63.28$\pm$09.18 & \textbf{67.27$\pm$09.18} \\
\midrule
\multicolumn{2}{l}{Average Performance} & 64.69$\pm$06.76 & 63.51$\pm$07.63 & 62.94$\pm$08.88 &  62.70$\pm$08.65 & \textbf{67.08$\pm$09.24} \\
\bottomrule
\end{tabular*}}
\end{center}
\end{table*}



 
%

\bibliographystyle{IEEEtran}
\bibliography{references}



 




\vfill

\end{document}